\documentclass[useAMS,usenatbib,usegraphicx,a4paper]{mn2e}

\newcommand{\apj}{ApJ} 
\newcommand{\aap}{A\&A} 
\newcommand{\mnras}{MNRAS} 

\title[HerMES: \textit{Herschel}--SPIRE point source catalogues]{HerMES: point source catalogues from deep \textit{Herschel}--SPIRE observations\thanks{\textit{Herschel} is an ESA space observatory with science instruments provided by European-led Principal Investigator consortia and with important participation from NASA.}}

\author[A.J.~Smith et al.]
{\parbox{\textwidth}{\raggedright \Large A.J.~Smith,$^{1}$\thanks{E-mail: \texttt{A.J.Smith@Sussex.ac.uk}}
L.~Wang,$^{1}$
S.J.~Oliver,$^{1}$
R.~Auld,$^{2}$
J.~Bock,$^{3,4}$
D.~Brisbin,$^{5}$
D.~Burgarella,$^{6}$
P.~Chanial,$^{7}$
E.~Chapin,$^{8}$
D.L.~Clements,$^{9}$
L.~Conversi,$^{10}$
A.~Cooray,$^{11,3}$
C.D.~Dowell,$^{3,4}$
S.~Eales,$^{2}$
D.~Farrah,$^{1}$
A.~Franceschini,$^{12}$
J.~Glenn,$^{13,14}$
M.~Griffin,$^{2}$
R.J.~Ivison,$^{15,16}$
A.M.J.~Mortier,$^{9}$
M.J.~Page,$^{17}$
A.~Papageorgiou,$^{2}$
C.P.~Pearson,$^{18,19}$
I.~P{\'e}rez-Fournon,$^{20,21}$
M.~Pohlen,$^{2}$
J.I.~Rawlings,$^{17}$
G.~Raymond,$^{2}$
G.~Rodighiero,$^{12}$
I.G.~Roseboom,$^{1,16}$
M.~Rowan-Robinson,$^{9}$
R.~Savage,$^{1,22}$
Douglas~Scott,$^{8}$
N.~Seymour,$^{23,17}$
M.~Symeonidis,$^{17}$
K.E.~Tugwell,$^{17}$
M.~Vaccari,$^{12}$
I.~Valtchanov,$^{10}$
L.~Vigroux,$^{24}$
R.~Ward,$^{1}$
G.~Wright$^{15}$ and
M.~Zemcov$^{3,4}$}\vspace{0.4cm}\\
\parbox{\textwidth}{\raggedright \tiny $^{1}$Astronomy Centre, Dept. of Physics \& Astronomy, University of Sussex, Brighton BN1 9QH, UK\\
$^{2}$School of Physics and Astronomy, Cardiff University, Queens Buildings, The Parade, Cardiff CF24 3AA, UK\\
$^{3}$California Institute of Technology, 1200 E. California Blvd., Pasadena, CA 91125, USA\\
$^{4}$Jet Propulsion Laboratory, 4800 Oak Grove Drive, Pasadena, CA 91109, USA\\
$^{5}$Department of Astronomy, Space Science Building, Cornell University, Ithaca, NY, 14853-6801, USA\\
$^{6}$Laboratoire d'Astrophysique de Marseille, OAMP, Universit\'e Aix-marseille, CNRS, 38 rue Fr\'ed\'eric Joliot-Curie, 13388 Marseille cedex 13, France\\
$^{7}$Laboratoire AIM-Paris-Saclay, CEA/DSM/Irfu - CNRS - Universit\'e Paris Diderot, CE-Saclay, pt courrier 131, F-91191 Gif-sur-Yvette, France\\
$^{8}$Department of Physics \& Astronomy, University of British Columbia, 6224 Agricultural Road, Vancouver, BC V6T~1Z1, Canada\\
$^{9}$Astrophysics Group, Imperial College London, Blackett Laboratory, Prince Consort Road, London SW7 2AZ, UK\\
$^{10}$Herschel Science Centre, European Space Astronomy Centre, Villanueva de la Ca\~nada, 28691 Madrid, Spain\\
$^{11}$Dept. of Physics \& Astronomy, University of California, Irvine, CA 92697, USA\\
$^{12}$Dipartimento di Astronomia, Universit\`{a} di Padova, vicolo Osservatorio, 3, 35122 Padova, Italy\\
$^{13}$Dept. of Astrophysical and Planetary Sciences, CASA 389-UCB, University of Colorado, Boulder, CO 80309, USA\\
$^{14}$Center for Astrophysics and Space Astronomy 389-UCB, University of Colorado, Boulder, CO 80309, USA\\
$^{15}$UK Astronomy Technology Centre, Royal Observatory, Blackford Hill, Edinburgh EH9 3HJ, UK\\
$^{16}$Institute for Astronomy, University of Edinburgh, Royal Observatory, Blackford Hill, Edinburgh EH9 3HJ, UK\\
$^{17}$Mullard Space Science Laboratory, University College London, Holmbury St. Mary, Dorking, Surrey RH5 6NT, UK\\
$^{18}$RAL Space, Rutherford Appleton Laboratory, Chilton, Didcot, Oxfordshire OX11 0QX, UK\\
$^{19}$Institute for Space Imaging Science, University of Lethbridge, Lethbridge, Alberta, T1K 3M4, Canada\\
$^{20}$Instituto de Astrof{\'\i}sica de Canarias (IAC), E-38200 La Laguna, Tenerife, Spain\\
$^{21}$Departamento de Astrof{\'\i}sica, Universidad de La Laguna (ULL), E-38205 La Laguna, Tenerife, Spain\\
$^{22}$Warwick Systems Biology Centre, Coventry House, University of Warwick, Coventry CV4 7AL, UK\\
$^{23}$CSIRO Astronomy \& Space Science, PO Box 76, Epping, NSW 1710, Australia\\
$^{24}$Institut d'Astrophysique de Paris, UMR 7095, CNRS, UPMC Univ. Paris 06, 98bis boulevard Arago, F-75014 Paris, France}}

\usepackage{amsmath}
\usepackage{amssymb}
\usepackage{subfigure}
\usepackage[colorlinks,citecolor=black,urlcolor=black,linkcolor=black]{hyperref}

\date{Accepted 2011 August 25. Received 2011 August 25; in original form 2011 January 17}

\pagerange{\pageref{firstpage}--\pageref{lastpage}} \pubyear{2011}

\begin{document}

\label{firstpage}

\maketitle

\begin{abstract}
We describe the generation of single-band point source catalogues from submillimetre \textit{Herschel}--SPIRE observations taken as part of the Science Demonstration Phase of the \textit{Herschel} Multi-tiered Extragalactic Survey (HerMES). Flux densities are found by means of peak-finding and the fitting of a Gaussian point-response function. With highly-confused images, careful checks must be made on the completeness and flux density accuracy of the detected sources. This is done by injecting artificial sources into the images and analysing the resulting catalogues. Measured flux densities at which 50 per cent of injected sources result in good detections at (250, 350, 500)\,$\mu$m range from (11.6, 13.2, 13.1)\,mJy to (25.7, 27.1, 35.8)\,mJy, depending on the depth of the observation (where a `good' detection is taken to be one with positional offset less than one full-width half-maximum of the point-response function, and with the measured flux density within a factor of 2 of the flux density of the injected source). This paper acts as a reference for the 2010 July HerMES public data release.
\end{abstract}

\begin{keywords}
catalogues -- submillimetre: galaxies -- methods: data analysis -- galaxies: photometry
\end{keywords}

\section{Introduction}

Since the discovery of the far-infrared background (FIRB; \citealt{Puget...1996,Fixsen...1998,Dwek...1998}), successive surveys have aimed to identify the discrete sources (primarily galaxies) responsible for this emission. With the launch of the ESA \textit{Herschel} Space Observatory \citep{Pilbratt...2010}, with its large (3.5\,m) telescope and high sensitivity, it is now possible to resolve a much greater fraction of the FIRB. An essential element of this is to have methods for identifying individual sources from \textit{Herschel} data.

This paper describes the generation of single-band point source catalogues from scan-map observations at 250, 350 and 500\,$\mu$m made using the photometer array of the SPIRE instrument on \textit{Herschel}. The SPIRE instrument, its in-orbit performance, and its scientific capabilities are described by \citet{Griffin...2010}, and the SPIRE astronomical calibration methods and accuracy are outlined by \citet{Swinyard...2010a}. The observations described here have been taken as part of the \textit{Herschel} Multi-tiered Extragalactic Survey (HerMES; Oliver et al., in preparation),\footnote{\url{http://hermes.sussex.ac.uk}} using data from the Science Demonstration Phase (SDP) of the survey. These observations cover approximately 20\,deg$^2$ in five regions located in four extra-Galactic fields, chosen for their minimal Galactic emission at far-infrared wavelengths, and for the amount of high-quality multi-wavelength ancillary data available in those fields (Oliver et al., in preparation).

Details of the observations are given in Table \ref{tbl:wcs}. The observations in the Spitzer First Look Survey (FLS) field were taken in SPIRE--PACS parallel mode, at scan speed 20 arcsec per second, while the other observations were taken in SPIRE-only mode, at scan speed 30 arcsec per second (Abell 2218, GOODS-North and Lockman-North) or 60 arcsec per second (Lockman-SWIRE). Standard SPIRE observing modes were used for all observations. The number of repetitions is indicated in Table \ref{tbl:wcs}; for each SPIRE-only repetition, the field is scanned in both the nominal and orthogonal directions, while for SPIRE--PACS parallel mode (FLS), one of the repetitions is in the nominal direction and the other is in the orthogonal direction. For the Lockman-SWIRE field, two separate observations were taken, offset from one another, in order to produce a more uniform coverage. The Abell 2218 (A2218) data were obtained through two observations, each consisting of 50 repetitions, separated by 38 days, giving complementary scan directions. All observations were taken with nominal bias mode. More details are given by \citeauthor{Oliver...2010} (\citeyear{Oliver...2010}; in preparation).

Subsets of some of the catalogues described here have been released to the public, as described in Appendix \ref{sec:release}.

\begin{table*}
\caption{\label{tbl:wcs}HerMES SDP SPIRE observations. For each field we give parameters for a rectangular region that avoids the edges of the fields, the total coverage of the observation being slightly larger. The roll angle is measured East of North for the shorter axis. $\langle N_{\rm samp}\rangle$ is the mean number of bolometer samples per pixel in the same typical-coverage region of the 250 $\mu $m map ($6 \times 6$ arcsec pixels). The number of repetitions is indicated, as described in the main text. Those fields for which a Wiener filter was applied to the map data are indicated (see Section \ref{sec:maps}).}
\vspace{0.2cm}
\centering
\begin{tabular}{l|rrrrrrrr}
\hline
\hline
Name 			& $N_\mathrm{rep}$	& RA /$^\circ$	& Dec /$^\circ$	& Roll /$^\circ$ & Size & $\langle N_{\rm samp}\rangle$ & Wiener filter \\
\hline
Abell 2218 (A2218)	& 100	& 248.98 		& 66.22 		& 37 			& $9^{\prime} \times 9^{\prime}$		& 1622	& N  \\ 
Spitzer First Look Survey (FLS)
  				& 2	& 258.97 		& 59.39 		& 5 			& $155^{\prime} \times 135^{\prime}$	& 30	& Y  \\ 
GOODS-North 		& 30	& 189.23 		& 62.24 		& 42	 		& $30^{\prime} \times 30^{\prime}$ 		& 501	& N  \\ 
Lockman-North 	& 7	& 161.50 		& 59.02 		& 1 			& $35^{\prime} \times 35^{\prime}$ 		& 117	& N  \\
Lockman-SWIRE 	& 2	& 162.00 		& 58.11 		& 2 			& $218^{\prime} \times 218^{\prime}$ 	& 16	& Y  \\
\hline
\hline
\end{tabular}
\end{table*}

This paper is organised as follows. Section \ref{sec:mapsandcats} describes the generation of maps and catalogues from the raw SPIRE data. Essential for any statistical analysis of a source catalogue is to quantify the completeness and reliability of the catalogue, and any systematic errors in the flux densities and positions. This is investigated in Section \ref{sec:metrics}, which describes a formalism for measuring these quantities and then applies that formalism to the catalogues. Conclusions are presented in Section \ref{sec:conclusions}.

\section{Catalogue generation}
\label{sec:mapsandcats}

The data processing occurs in several distinct stages, each of which is described here.

\subsection{Timelines}
\label{sec:timelines}

The SPIRE photometer contains three bolometer arrays observing simultaneously at 250, 350 and 500\,$\mu$m. The observations were taken as scan maps, with the telescope scanning the survey region at a constant rate, and with the voltage across each of the bolometers in the three SPIRE arrays being sampled at least 10 times per second (specifically, 18.6\,Hz for SPIRE-only observations and 10\,Hz for observations taken in SPIRE-PACS parallel mode, \citealt{Griffin...2010}). For each scan leg, this results in a series of samples for each bolometer, known as a `timeline'.

The raw timelines were processed using the standard SPIRE photometer pipeline \citep{Dowell...2010} to produce calibrated and corrected timelines in units of Jy. Specifically, the pipeline used was that provided in \textsc{hipe} \citep{Ott2010} development version 2.0.905, with a fix applied to correct for a gradual drift in the astrometry (included in more recent versions of the pipeline), and using the following calibration products: beam-steering mirror calibration version 2, flux conversion version 2.3 and temperature drift correction version 2.3.2.

A small number of cosmic ray hits (`glitches') were not detected by the pipeline and were propagated through to the maps; see Section \ref{sec:reliability} for a discussion of the effects this has on the final catalogues.

A multiplicative correction was applied to the pipeline flux densities of (1.0, 1.02, 0.92) for (250, 350, 500)\,$\mu$m. These factors were the best estimate of the correction factors at the time the data were processed; subsequent analysis measured the correction factors to be (1.02, 1.05, 0.94), as given by \citet{Griffin...2010}. The current photometric accuracy of SPIRE, based on Ceres observations and models, is estimated to be 15 per cent \citep{Swinyard...2010a} at each band, with a high correlation between bands.

\subsection{Maps}
\label{sec:maps}

From the timelines, maps were created using the default \textsc{hipe} naive map-maker, with the default pixel sizes of (6, 10, 14) arcsec for (250, 350, 500)\,$\mu$m. In map pixel $i$, the signal, $d_i$, is estimated from the $N_i$ bolometer samples $\{ s_j \}$ lying within that pixel as
\begin{eqnarray}
\label{eqn:mapflux}
d_i = \bar{s} = \frac{1}{N_i} \sum_{j=1}^{N_i} s_j
\end{eqnarray}
while $\sigma_i$, the uncertainty in the value of $d_i$, is the standard error of the mean for $\{s_j\}$:
\begin{eqnarray}
\label{mapfluxerror}
\sigma_i = \left[ \frac{1}{N_i(N_i-1)} \sum_{j=1}^{N_i} (s_j - \bar{s})^2 \right]^{1/2}
\end{eqnarray}
Prior to map-making, the residual drift present in the timelines, which is a residual from the temperature drift correction \citep{Griffin...2010}, was removed by fitting a constant plus a linear slope to each scan timeline. Another (small) offset was then applied to give the maps a mean value of zero, since the true (physical) zero-point for the maps is unknown.

The overall astrometry of the maps has been adjusted by comparison with known positions of radio sources.  This has typically been a correction of around 2 or 3 arcsec, which is consistent with the absolute pointing error of \textit{Herschel} \citep{Pilbratt...2010}. 

Maps have been created using all of the data for each field, and also using two halves of the data, separated in time, to create two independent maps of each field, useful for confirmation and reliability purposes. (Note that for FLS one of these maps has data taken in the nominal direction and the other has data taken in the orthogonal direction; for all other fields both maps contain cross-scan data.)

\begin{figure}
\centering
\includegraphics[width=0.45\textwidth]{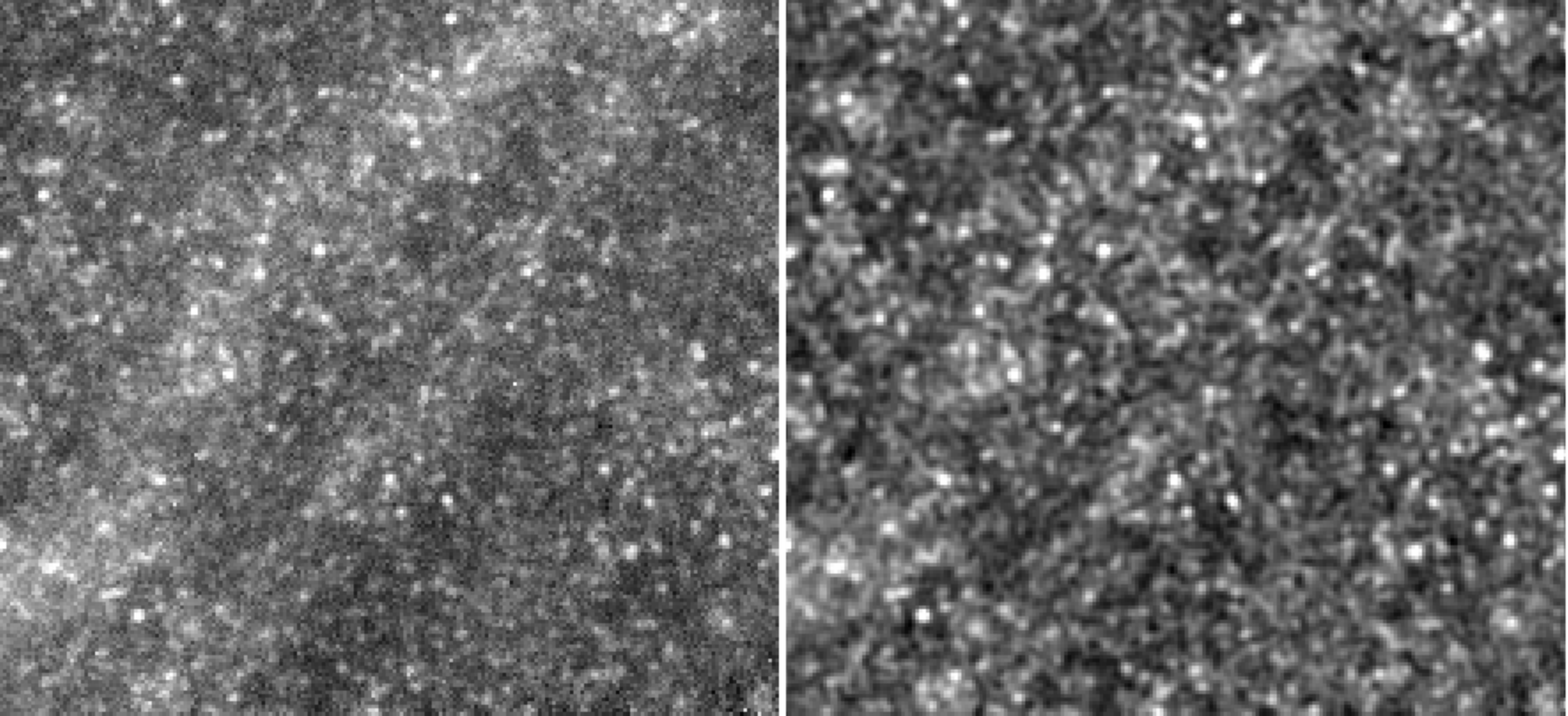}
\caption{\label{FLS_wiener}Part of the 350\,$\mu$m map from FLS, covering an area of approximately 36 arcmin by 33 arcmin, before (left) and after (right) applying the Wiener filter. The point sources are significantly emphasised by the Wiener filter, thus diminishing the effects of the large-scale diffuse cirrus.}
\end{figure}

For the shallowest fields (Lockman-SWIRE and FLS), there was diffuse cirrus clearly visible in parts of the maps. In order to accentuate the signal from point sources, and thus to reduce the effects of the cirrus, these maps have been modified using a Wiener filter \citep{Wiener1949,WallJ2003}, which is given by 
\begin{eqnarray}
F(f) = \frac{ |S(f)|^2 }{  |S(f)|^2 + |N(f)|^2 }\,,
\end{eqnarray}
where $f$ is the frequency, $S$ is the signal spectrum and $N$ the noise spectrum. The model for the signal is obtained from a noiseless simulation of sources with BLAST number counts \citep{Patanchon...2009}, which are in good agreement with the number counts estimated from these data \citep{Oliver...2010}. The model for the noise is obtained from the difference map of the two independent maps of each field (which gives approximately white noise). The absolute calibration of the filtered maps is not determined at this stage; instead, the flux densities measured are adjusted by injecting artificial sources into the map before applying the filter (see Section \ref{sec:completeness}). Fig.\ \ref{FLS_wiener} shows part of the FLS 350\,$\mu$m map before and after the Wiener filter has been applied.

For these same fields, a small number of individual scans would have produced obvious artefacts in the final maps and were therefore removed. (This was due to a combination of steps in the thermistor timelines and the temperature drift correction used.) In these regions, the coverage has consequently been reduced by approximately 25 per cent. For FLS, out of 117 scans, this has affected two scans obtained at 250\,$\mu$m and one scan obtained at 500\,$\mu$m, while for Lockman-SWIRE, out of 160 scans, the number affected at (250, 350, 500)\,$\mu$m was (3, 1, 2), respectively.

\subsection{Catalogues}
\label{sec:catalogues}

Source catalogues have been generated for each band in each field. Details of the columns are given in Appendix \ref{sec:release}. Source flux densities have been estimated using the \textsc{sussextractor} point source extractor \citep{SavageO2007} as implemented in \textsc{hipe} 3.0.  For computational efficiency, a Gaussian point-response function (PRF) was assumed, with full-width half-maximum (FWHM) of (18.15, 25.15, 36.3) arcsec for (250, 350, 500)\,$\mu$m, and with a Gaussian approximation used for the beam area:
\begin{eqnarray}
\Omega = \frac{\pi(\text{FWHM})^2}{4\ln 2}\,.
\end{eqnarray}
\citet{Griffin...2010} state that Gaussian beams with FWHM of (18.1, 25.2, 36.6) arcsec provide a good approximation to the true beam, so we assume the errors introduced by our choice of PRF will be small compared with other sources of uncertainty.

The flux density is given by
\begin{eqnarray}
\label{eqn:flux}
S = \sum_{i=1}^{N_\mathrm{pixels}} \frac{d_i \mathcal{P}_i}{\sigma_i^2} \Bigg/ \sum_{i=1}^{N_\mathrm{pixels}} \frac{\mathcal{P}_i^2}{\sigma_i^2} \,,
\end{eqnarray}
where the summation is over a local region around the source position, $d_i$ is the value of the map pixel, $\sigma_i$ is the value of the error map pixel and $\mathcal{P}_i$ is the smoothing kernel (matched filter). The uncertainty in the flux density is discussed in Section \ref{sec:uncertainties}.

For isolated point sources and white noise, the optimal matched filter is the beam itself. But for a higher density of sources the optimal filter will be narrower than the beam, up to the limit of complete confusion (no white noise), in which the optimal filter is a deconvolution filter, given by the inverse of the beam \citep[see][appendix A]{Chapin...2010}.

However, the method used here, for convenience, was to take as the smoothing kernel the central region of a Gaussian PRF centred on the pixel closest to the source position. \textsc{sussextractor} was used in two distinct ways, one for the shallower fields, and one for the deeper, so as to deal with the high pixel-to-pixel noise in the shallower maps and to exploit the high signal-to-noise in the deeper maps.

For the shallower fields (FLS, Lockman-SWIRE and Lockman-North), a single pass was performed, which involves \textsc{sussextractor} applying a smoothing kernel to the image (the central $5 \times 5$ pixels of the Gaussian PRF), to obtain an image in which the value of each pixel is the maximum likelihood estimate of the flux density of a source centred on the centre of that pixel, as given in Equation (\ref{eqn:flux}). In this smoothed map, the extractor then searched for local maxima, comparing each pixel with its eight immediate neighbours, and the value at these peak positions was taken as the estimate of the source flux density. The position of the source was refined to sub-pixel accuracy based on the intensity of the surrounding pixels.

For the deeper fields (A2218 and GOODS-North), first \textsc{sussextractor} was run using no smoothing, in order to find the positions of the local maxima in the image. Then \textsc{sussextractor} was run again, this time with a small smoothing kernel (the central $3 \times 3$ pixels of the Gaussian PRF), in order to estimate the source flux densities at these position. This method has been adopted in order to extract to fainter flux densities, and in order to reduce the number of close pairs of sources that are blended into a single source in the catalogue.

Sources have been extracted close to the edge of the images. However, a central region has been defined for each field, so that sources can be selected within a simple rectangular region of the image, avoiding the edges. Parameters defining these regions are in Table \ref{tbl:wcs}. These regions have an easily-determined area, and therefore the subsets can be used for studies of the number density of sources.

At the positions of the sources in the catalogues, flux densities have been estimated from the two independent maps, each produced from half of the data (see Section \ref{sec:maps}). These flux densities have been included in the catalogues, and may be used for investigations of the reliability of sources (see Section \ref{sec:reliability}).

A multiplicative factor has been applied to all flux densities, in order to give approximately zero mean offset in log-flux density for the brightest injected sources (see Section \ref{sec:completeness_results}). This is to account for the arbitrary normalisation and other effects of the Wiener filter, and also for a systematic underestimation of the flux densities by \textsc{sussextractor}, primarily due to the assumption in the flux density estimation that the source centre is aligned with the centre of a pixel. First, some bright sources of flux density $S$ were added to the images. Then the flux densities of these sources were measured. If the mean measured flux density was $S_\mathrm{mean}$, then the multiplicative factor was chosen to be $S / S_\mathrm{mean}$. For the (250, 350, 500)\,$\mu$m flux densities, the factors applied to the standard (naive map) data were A2218: (1.052, 1.062, 1.040), GOODS-North: (1.067, 1.062, 1.074), Lockman-North: (1.028, 1.040, 1.038). For FLS and Lockman-SWIRE, the factors applied to the Wiener-filtered data were (1.266, 1.304, 1.482) for FLS and (1.613, 1.580, 1.713) for Lockman-SWIRE. (Note that these factors were derived before certain improvements were made to the method of injecting artificial sources, so the offset in log-flux density is only approximately zero for bright injected sources.)

\subsection{Uncertainties in source flux densities}
\label{sec:uncertainties}

The formal uncertainty in the flux density in Equation (\ref{eqn:flux}) is given by
\begin{eqnarray}
\label{eqn:fluxerror}
\sigma_\mathrm{S} = 1 \Bigg/ \sqrt{\sum_{i=1}^{N_\mathrm{pixels}} \frac{\mathcal{P}_i^2}{\sigma_i^2}}
\end{eqnarray}
It should be noted that in expressing the uncertainty in the flux in this way, it has been assumed implicitly that the covariance between pixels is negligible. For the naive maps, this is a reasonable assumption, but for the Wiener-filtered data (FLS and Lockman-SWIRE) the covariance between neighbouring pixels is not negligible, and thus the formal uncertainty in the flux density from Equation (\ref{eqn:fluxerror}) will be a poor estimate of the formal uncertainty. Moreover, for the Wiener-filtered data, the error map (which was not Wiener filtered) has also been scaled by the (large) multiplicative factors given in Section \ref{sec:catalogues}, which will significantly increase the value of $\sigma_S$. These values should therefore not be over-interpreted for the Wiener-filtered data. However, the total noise estimates (see below) are based primarily on the statistics of the smoothed images provided by the source extraction software, rather than on the formal uncertainty in Equation (\ref{eqn:fluxerror}), so these are more robust against these effects.

These flux density uncertainties are believed (for the non-Wiener-filtered data) to give a fair estimate of the \textit{instrumental} noise, and will be referred to as such hereafter. But they do not include the effects of source confusion (the high density of sources relative to the size of the SPIRE beams) nor the effects of the uncertainty in the PRF or the SPIRE absolute flux calibration. So the true uncertainty in the source flux density will be significantly higher than the instrumental noise.

The \textit{total} noise, taking account of confusion noise as well as instrumental noise, is estimated as follows:
\begin{enumerate}
\item The smoothed map is obtained from the source extraction software, with the value in each pixel, $j$, being an estimate of the flux density, $S_j$, of a point source assumed to lie at the centre of that pixel, calculated from Equation (\ref{eqn:flux}).
\item The typical total noise in the source flux density, $\sigma_\mathrm{total}$, is derived from the statistics of this smoothed map:
\begin{eqnarray}
\sigma_\mathrm{total} = \sqrt{\frac{\sum_j \left( S_j - \text{median}(S_j) \right)^2 }{N}} \,,
\end{eqnarray}
where the summation is over the central region of the map, restricted to those pixels in which $S_j < \text{median}(S_j)$, and where $N$ is the number of pixels included in the summation.
\item The typical instrumental noise, $\sigma_\mathrm{instrumental}$, is estimated as the median of Equation (\ref{eqn:fluxerror}), over the central region of the map:
\begin{eqnarray}
\sigma_\mathrm{instrumental} = \text{median} \left( \sigma_{\mathrm{S},j} \right) \,.
\end{eqnarray}
\item A single value for the typical contribution from source confusion to the uncertainty in the flux density of a point source is derived by subtracting, in quadrature, the typical instrumental noise from the typical total noise:
\begin{eqnarray}
\label{eqn:confusion}
\sigma_\mathrm{confusion}^2 = \sigma_\mathrm{total}^2 - \sigma_\mathrm{instrumental}^2\,.
\end{eqnarray}
This is a similar, but not identical, quantity to the `confusion noise' derived by \citet{Nguyen...2010}, which is the contribution from source confusion to the uncertainty in the intensity of a typical map pixel and is measured to be `5.8, 6.3 and 6.8 mJy/beam at 250, 350 and 500 $\mu$m, respectively'.
\item The total noise for each individual source (pixel $j$) is then obtained by adding, in quadrature, this typical value for the confusion noise to the source's own instrumental noise value, given by Equation (\ref{eqn:fluxerror}):
\begin{eqnarray}
\sigma_{\mathrm{instrumental},j}^2 = \sigma_\mathrm{confusion}^2 + \sigma_{\mathrm{S},j}^2 \,.
\end{eqnarray}
\end{enumerate}

The values for the typical total noise and the typical instrumental noise are shown in Table \ref{tbl:noise}. Note that smoothing of the maps increases the confusion noise (although it decreases the instrumental noise contribution), since smoothing increases the size of the effective beam. More smoothing has been applied to the shallower fields (FLS, Lockman-North, Lockman-SWIRE) than to the deeper fields (A2218, GOODS-North), so greater confusion noise is to be expected in those fields. Moreover, the FLS and Lockman-SWIRE fields have been smoothed using a Wiener filter (with different filters for each field), which will broaden the PRF and thus increase the confusion further.

As discussed above, the measurement of the instrumental noise for the Wiener-filtered data (FLS and Lockman-SWIRE) is believed to be over-estimated, so it is not possible to give a reliable estimate of the confusion noise using Equation (\ref{eqn:confusion}). However, for the remaining data, $\sigma_\mathrm{confusion}$ is found to be (5.9, 7.5, 7.7)\,mJy for A2218, (5.6, 7.4, 7.7)\,mJy for GOODS-North and (6.8, 8.3, 8.5)\,mJy for Lockman-North, all for (250, 350, 500)\,$\mu$m respectively.

\addtocounter{footnote}{1}
\footnotetext[\value{footnote}]{Estimates of the noise for FLS and Lockman-SWIRE were revised slightly since the publication of \citet{Schulz...2010}, but the findings of that paper are unaffected.}

\begin{table}
\caption{\label{tbl:noise}Approximated $1\sigma$ uncertainty in the flux density of a typical point source, in mJy, from the combined effects of instrumental and confusion noise, as described in the text. Shown in parentheses is the median $1\sigma$ instrumental noise in the flux density measurement of a point source, in mJy. For the Wiener-filtered data (FLS and Lockman-SWIRE), the instrumental noise is believed to be over-estimated, and is shown in italics.}
\vspace{0.2cm}
\centering
\begin{tabular}{l|rrrrrr}
\hline
\hline
Field 	& \multicolumn{6}{c}{$\sigma_\mathrm{total}$ ($\sigma_\mathrm{instrumental}$), mJy} \\
	 	& \multicolumn{2}{c}{250\,$\mu$m} 	& \multicolumn{2}{c}{350\,$\mu$m} 	& \multicolumn{2}{c}{500\,$\mu$m} \\
\hline
A2218 			& 5.9 	& (0.6)		& 7.6 		& (0.6)		& 7.8 	& (0.6) \\
FLS\footnotemark[\value{footnote}]
	    			& 8.8 	& \textit{(3.1)}	& 10.0 		& \textit{(3.2)}	& 11.1 	& \textit{(4.5)}  \\
GOODS-North		& 5.7 	& (0.9) 		& 7.4 		& (0.9)		& 7.8 	& (1.1) \\
Lock.-North		& 7.0 	& (1.7)		& 8.5 		& (1.7)		& 8.8 	& (2.1) \\
Lock.-SWIRE\footnotemark[\value{footnote}]
				& 10.4 	& \textit{(6.6)}	& 11.6 		& \textit{(6.5)}	& 11.8 	& \textit{(8.7)} \\
\hline
\hline
\end{tabular}
\end{table}

The initial threshold on the catalogues is $3\sigma$, based on the instrumental noise. Some further cuts have been applied to the released catalogues (see Appendix \ref{sec:release}).
 
\begin{table*}
\caption{\label{tbl:counts}Number of sources in the central region of each field with signal-to-(total) noise greater than 3. In parentheses are the number of sources with signal-to-(instrumental) noise greater than 3.}
\vspace{0.2cm}
\centering
\begin{tabular}{l|rrrrrr}
\hline
\hline
Field 	& \multicolumn{6}{c}{Number of sources $>$ 3$\sigma_\mathrm{total}$ ($>$ 3$\sigma_\mathrm{instrumental}$)} \\
	 	& \multicolumn{2}{c}{250\,$\mu$m} 	& \multicolumn{2}{c}{350\,$\mu$m} 	& \multicolumn{2}{c}{500\,$\mu$m} \\
\hline
A2218 			& 41		& (119)		& 12		& (64)	& 5		& (36) 	\\
FLS	  			& 3946  	& (12\,862) 	& 1822 	& (7120) 	& 637 	& (2751)	\\
GOODS-North		& 385	& (1421)		& 150  	& (713) 	& 48 		& (344) 	\\
Lockman-North		& 325	& (1082) 	  	& 141  	& (586) 	& 61 		& (255)	\\
Lockman-SWIRE	& 6731	& (13\,890) 	& 2757	& (7867) 	& 836	& (1902) 	\\
\hline
\hline
\end{tabular}
\end{table*}

\begin{figure*}
\centering
\subfigure[A2218]{\includegraphics[width=.45\textwidth]{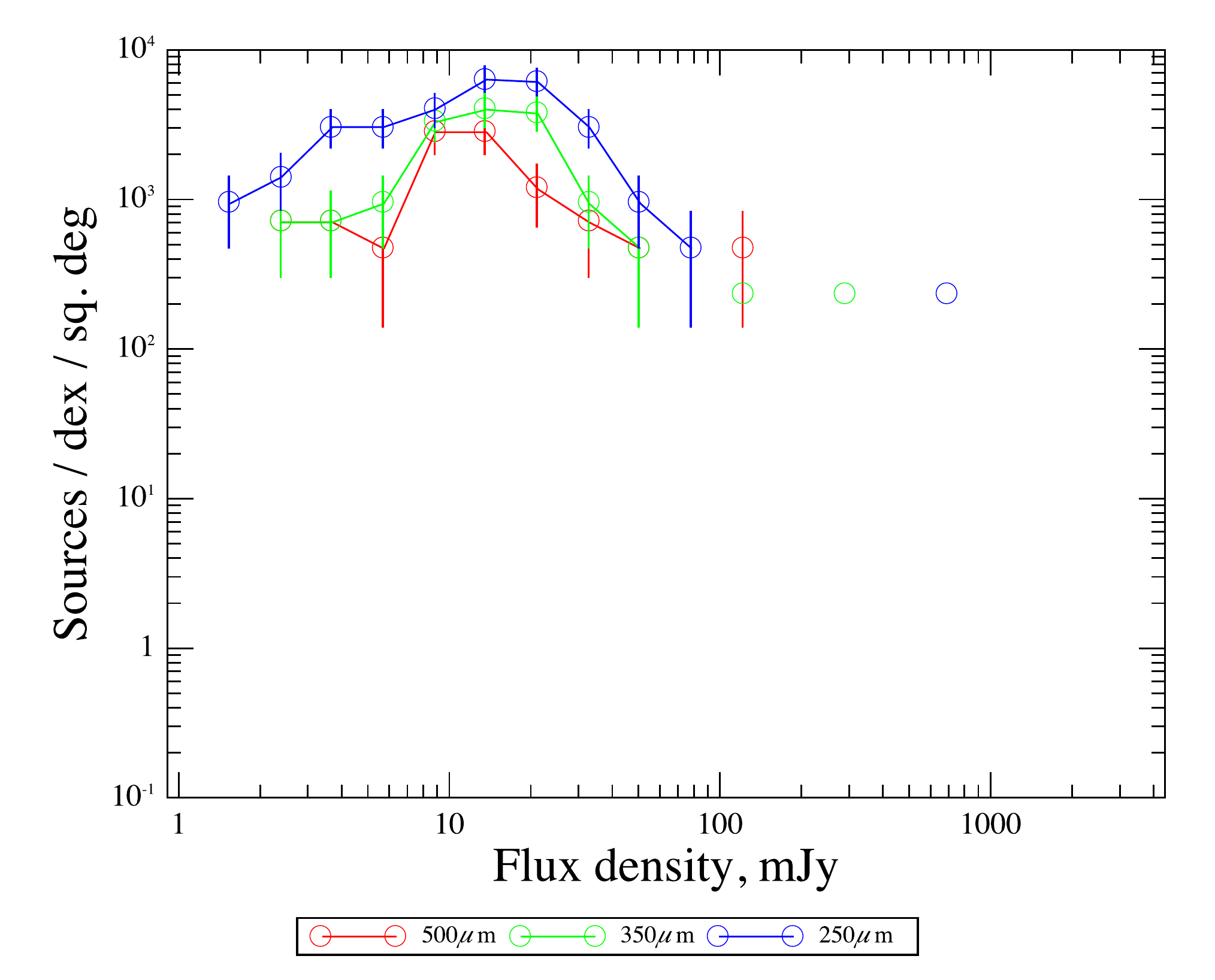}}
\subfigure[FLS]{\includegraphics[width=.45\textwidth]{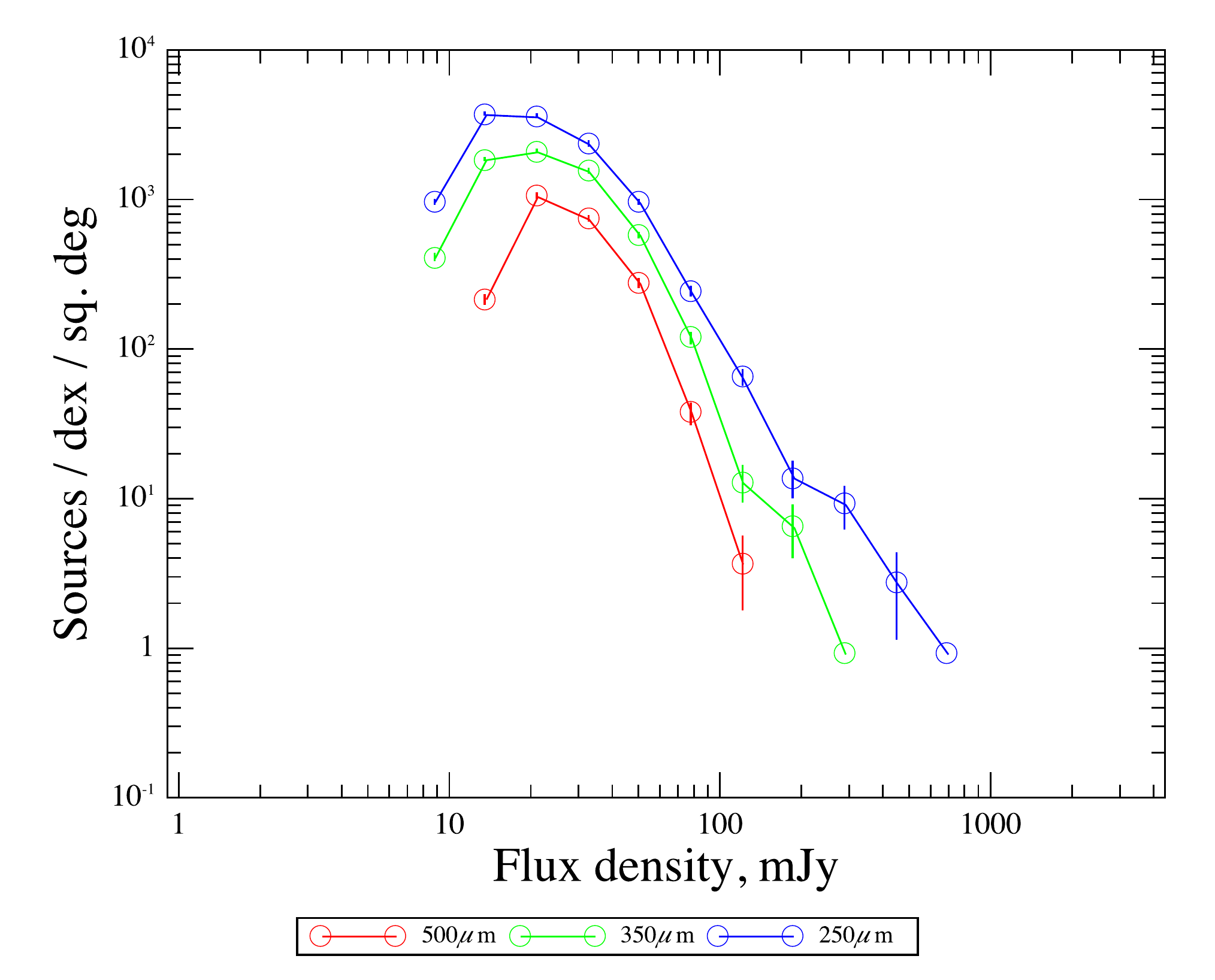}}
\subfigure[GOODS-North]{\includegraphics[width=.45\textwidth]{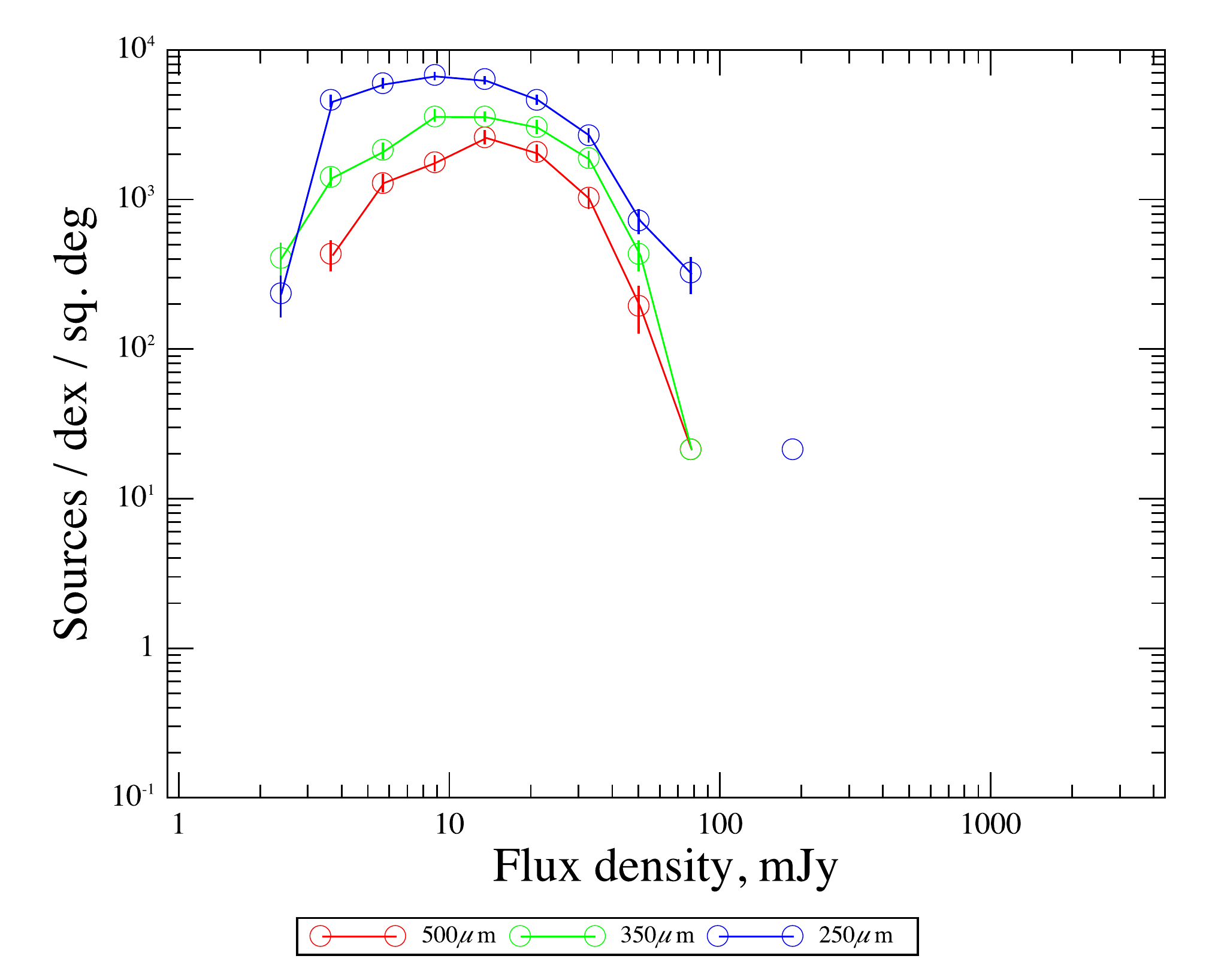}}
\subfigure[Lockman-North]{\includegraphics[width=.45\textwidth]{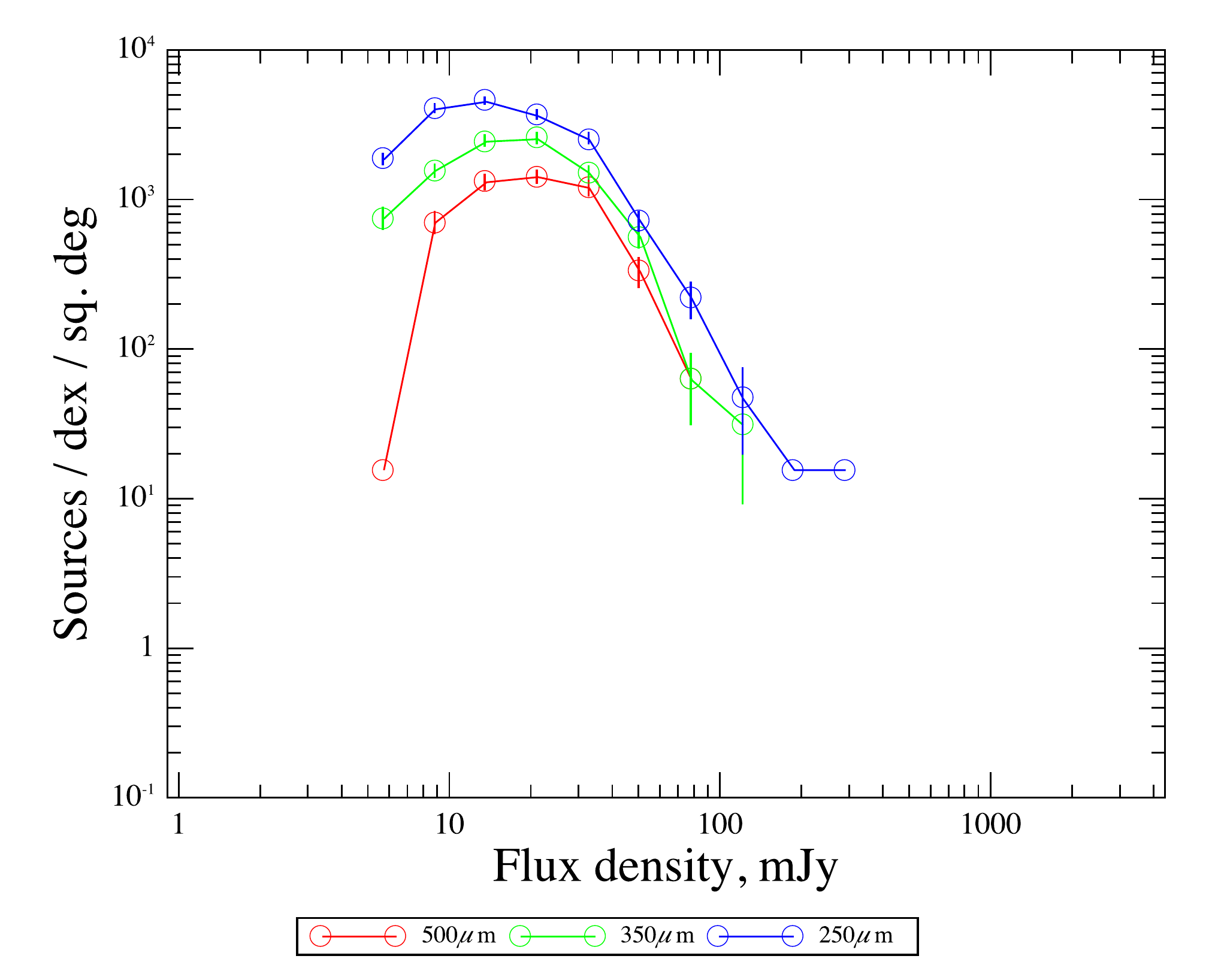}}
\subfigure[Lockman-SWIRE]{\includegraphics[width=.45\textwidth]{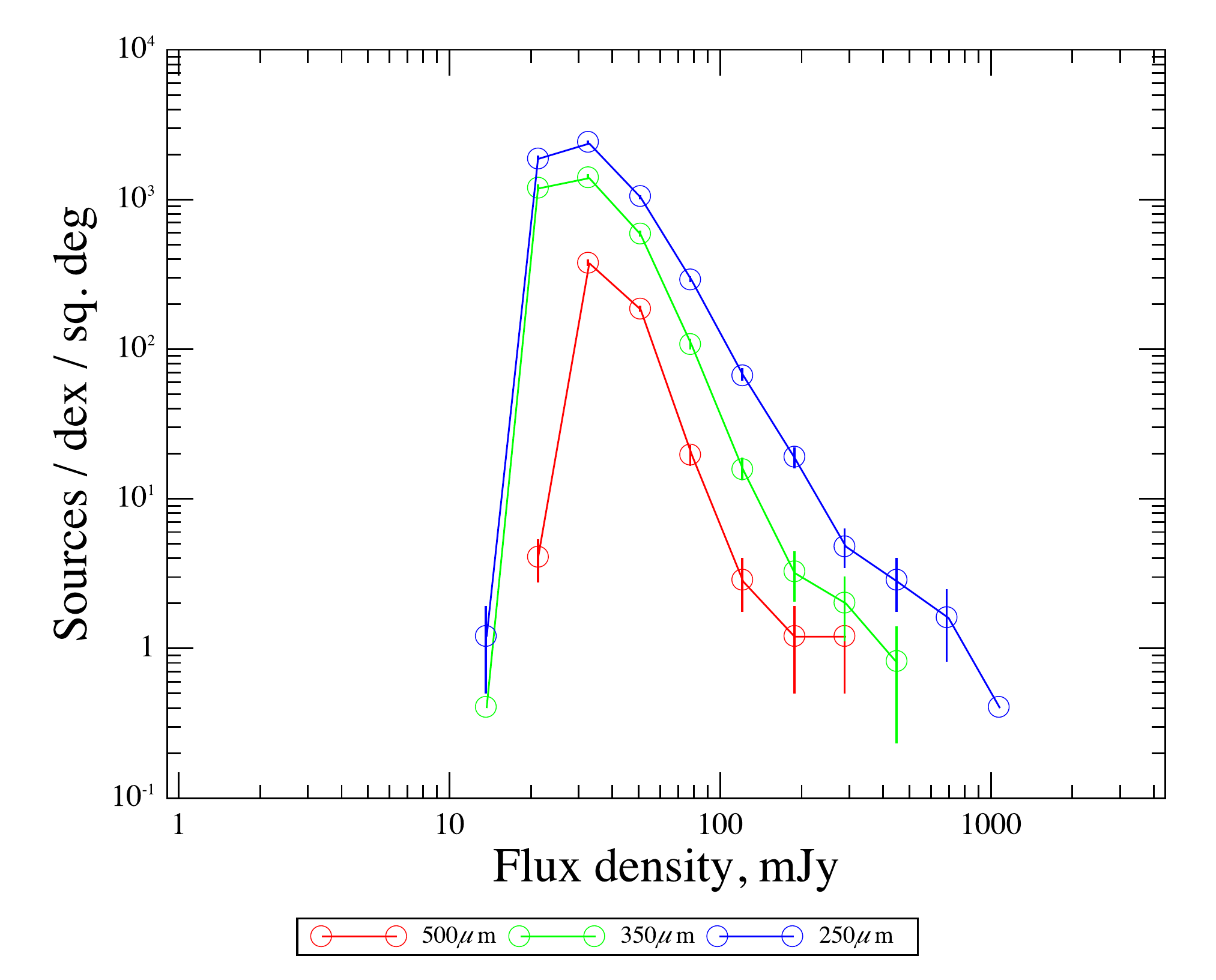}}
\caption{\label{fig:counts}Raw source counts for the five fields, showing the number of sources per unit area per $\log_{10}$ interval in the central region. No corrections have been applied for incompleteness or flux boosting. Error bars are the counts in that bin divided by the square root of the number of sources in the bin. Each plot shows the results for 250\,$\mu$m (blue), 350\,$\mu$m (green) and 500\,$\mu$m (red).}
\end{figure*}

The raw source counts for the central region of each field are shown in Fig.\ \ref{fig:counts} with the total number of sources in the central region of each field, and the number with signal-to-(total) noise ratio greater than 3, shown in Table \ref{tbl:counts}.

\section{Quality of the catalogues}
\label{sec:metrics}

A catalogue will be of limited use without some measure or assurance of its quality. This may be the reliability (number of false detections), completeness (probability that a genuine source will be included in the catalogue) or the accuracy of the parameters of the sources (position and flux density). These will be discussed below.

\subsection{Reliability}
\label{sec:reliability}

The reliability of a source catalogue is conventionally a measure of the fraction of detections, at a given flux density, that are spurious. A spurious detection may happen as a result of noise in the map pixels (due to a small number of bolometer samples, each with a significant uncertainty), or as a result of other factors contributing to the detector signal, such as any cosmic ray hits (glitches) that are not removed by the pipeline.

When the noise in the data is due entirely to these (instrumental) effects, the probability that a detection is genuine (or spurious) can be estimated from the signal-to-noise ratio of the source. However, in these \textit{Herschel}--SPIRE data, the dominant source of noise is confusion, that is, the measurement of the flux density of any particular source being contaminated by the flux density of neighbouring sources. This means that the signal-to-(total) noise of a detection cannot be used in any straightforward way to give the probability that it is spurious.

The number of such spurious detections that would arise from instrumental noise may be estimated using the maps and catalogues generated from the two halves of the data for each field (see Section \ref{sec:maps}).

If $d_1$ is the measured intensity in a map pixel from the first half of the data and $d_2$ is the measured intensity in the same pixel from the second half, then the intensity for the total map will be
\begin{eqnarray}
d_\mathrm{total} = \frac{d_1 + d_2}{2}\,.
\end{eqnarray}
Two `difference maps' may be obtained by taking the difference between these two measurements:
\begin{eqnarray}
\label{eqn:diff}
d_\mathrm{difference} = \pm \left( \frac{d_1 - d_2}{2} \right) \,.
\end{eqnarray}
This difference map is then an instrumental-noise map with astronomical flux (and confusion noise) removed. Executing the source extraction on these maps will give an estimate of the number of spurious detections that might be expected, in the absence of confusion noise.

Any unremoved cosmic ray hits will either leave a positive spike or a negative spike in the difference map, depending on which half of the data is affected. The source extraction is therefore executed on both the `positive' and `negative' difference maps, from Equation (\ref{eqn:diff}).

\begin{table*}
\caption{\label{tbl:diffCounts}Number of detections in the central regions of the difference maps greater than 3$\sigma_\mathrm{total}$. Each pair of numbers is the number of detections in the first and then the second difference map, corresponding to the positive and negative forms of Equation (\ref{eqn:diff}), respectively. In parentheses are the number of detections with flux density greater than 3$\sigma_\mathrm{instrumental}$. The numbers should be compared with those given in Table \ref{tbl:counts}.}
\vspace{0.2cm}
\centering
\begin{tabular}{lrrrrrr}
\hline
\hline
Field 	& \multicolumn{6}{c}{Sources $>$ 3$\sigma_\mathrm{total}$ ($>$ 3$\sigma_\mathrm{instrumental}$)} \\
	 	& \multicolumn{2}{c}{250\,$\mu$m} 	& \multicolumn{2}{c}{350\,$\mu$m} 	& \multicolumn{2}{c}{500\,$\mu$m} \\
\hline
A2218 			& 0+0	& (93+108)	& 0+0	& (62+66)		& 0+0	& (35+23)	\\
FLS	   	 		& 7+3	& (28+25)		& 4+4	& (11+10)		& 0+1	& (4+6)	\\
GOODS-North		& 0+0	& (2+1) 		& 0+0	& (3+0)		& 0+0	& (1+0)	\\
Lockman-North		& 0+0	& (9+6) 		& 0+0	& (5+7)		& 0+0	& (5+36)	\\
Lockman-SWIRE	& 2+6	& (3+7) 		& 1+1	& (1+1)		& 4+3	& (4+4)	\\
\hline
\hline
\end{tabular}
\end{table*}

Table \ref{tbl:diffCounts} shows the numbers of sources detected from these difference maps with additional details given below. The numbers of detections should be compared with those in Table \ref{tbl:counts}.

For A2218 (0.022\,deg$^2$), all of the `detections' in the difference maps have flux densities below 7.2\,mJy. The detections are generally found along stripes in the map that remain as a result of our method for subtracting baselines from the timelines (Section \ref{sec:timelines}).  A small number of detections in the difference maps are associated with bright sources: these may have arisen as a result of the strong gradients in the signal associated with the sides of the beam, or as a result of the ellipticity of the beam (7--12 per cent, \mbox{\citealt{Griffin...2010}}) and the change in position angle between the two observations.

For FLS (5.8\,deg$^2$), 7+25 of the `detections' at 250\,$\mu$m, 5+9 at 350\,$\mu$m and 0+6 at 500\,$\mu$m are above the $1\sigma$ total noise values in Table \ref{tbl:noise}, with one of these having a flux density greater than 200\,mJy. For Lockman-SWIRE (13\,deg$^2$), the measured flux densities are between 28 and 400\,mJy. By inspection of the images, all of these detections were found to be due to glitches that were not removed by the pipeline, with the exception of 12 detections in FLS at 250\,$\mu$m: 4 associated with bright sources, and 8 detections fainter than 10.6\,mJy.

For GOODS-North (0.25\,deg$^2$) and for Lockman-North (0.34\,deg$^2$), all of the measured flux densities are below $7.5$\,mJy, with the exception of the second difference map for Lockman-North at 500\,$\mu$m, which gives 33 detections along one scan line, due to residuals in the baseline subtraction, having flux densities between 6.3 and 12.7\,mJy.

The spurious detections due to unremoved glitches cause a detection in the map from one half of the data but not in the map from the other half. The flux densities generated from these half-data maps are included in the catalogues, and may be used to identify some of these.

\subsection{Completeness, flux density accuracy and positional accuracy: method}
\label{sec:completeness}

The completeness, flux density accuracy and positional accuracy have been investigated by injecting artificial sources into the timelines, and then creating new maps from those modified timelines. This ensures that both the signal and its uncertainty are modified by the injected sources, as given by Equations (\ref{eqn:mapflux}) and (\ref{mapfluxerror}). For the Wiener-filtered data, the error maps were created in this way, but the signal maps were created by injecting sources directly into the unfiltered images, and then applying the Wiener filter to the images.

\begin{figure}
\centering
\includegraphics[width=0.3\textwidth]{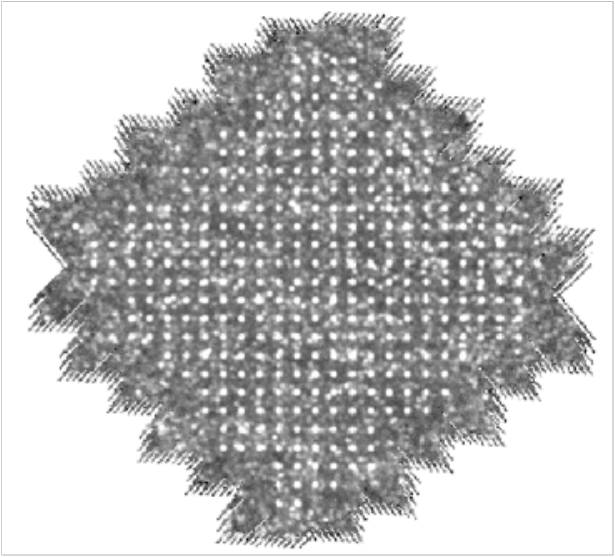}
\caption{\label{map_with_grid}The 350\,$\mu$m map from GOODS-North, with a grid of 100 mJy sources injected.}
\end{figure}

For each field, this is done multiple times, with the same flux density for all of the injected sources. Sources are placed on a grid, with spacing offset from the pixel size of the image and large enough that the sources can be treated independently (as an approximation to the process of adding sources one at at time at random positions). (The precise spacing chosen was 113.387 arcsec for A2218, GOODS-North and Lockman-North, increased by a factor of 3 for FLS and Lockman-SWIRE.) An example is shown in Fig.\ \ref{map_with_grid}.

The procedure for measuring the completeness and flux density accuracy is as follows.

First the source catalogues are produced:
\begin{enumerate}
\item The source extraction is performed on the map with no artificial sources added, as described in Section \ref{sec:catalogues}, in order to define a \textbf{reference catalogue} for each band.
\item A \textbf{truth catalogue} is created, consisting of the grid of artificial sources, and this is used to create maps with injected sources, with all such sources having the same flux density. The whole procedure is repeated with each iteration having a different flux density for the injected sources. The flux densities chosen are 1, 2, 3, 4, 7, 10, 20, 30, 40, 70, 100, 200, 300, 400, 700, 1000 and 4000\,mJy.
\item For each map with injected sources, the source extraction is performed, in exactly the same way as for the reference catalogue, to produce additional \textbf{source catalogues}, to be compared with the reference catalogues.
\end{enumerate}

Next, with the reference catalogue, truth catalogue of artificial sources, and source catalogues for each injected flux density, the catalogues are compared as follows, for each band and for each injected flux density:
\begin{enumerate}
\item The artificial source truth catalogue is first cross-matched with the reference catalogue from the real data. For each source in the truth catalogue, the closest match within 1 times the FWHM is chosen (if such a source is present). If this match has a flux density within a factor of 2 of the injected source flux density, the match is identified as a `good' match. Any such `good' matches are discarded from further analysis; otherwise, when these (serendipitous) matches are included, the measured completeness can be misleading, particularly for source extraction methods that produce a large number of spurious, faint detections. However, excluding these sources will have a small effect on the estimates of the completeness, because part of the incompleteness comes from the fact that sources can be too close to other sources and therefore not counted.
\item Next the truth catalogue (without the serendipitous sources from the previous step) is cross-matched with the source catalogue derived from the map with injected sources. `Good' matches are found, as above. The completeness is defined as the number of good matches divided by the number of injected sources (minus the serendipitous sources).

For example, if 200 sources with flux density 30 mJy are injected into the map, but 20 of those already (by chance) have `good' counterparts in the original map (without injected sources), then the remaining number of sources is 180. If 162 of these have good matches in the source list extracted from the map with injected sources, then the completeness at 30 mJy is $162 / 180 = 90$ per cent.
\item The flux density and positional accuracy are found by comparing the extracted flux densities and positions with the injected flux densities and positions.
\end{enumerate}

\subsection{Completeness, flux density accuracy and positional accuracy: results}
\label{sec:completeness_results}

\begin{figure*}
\centering
\subfigure[A2218]{\includegraphics[width=.45\textwidth]{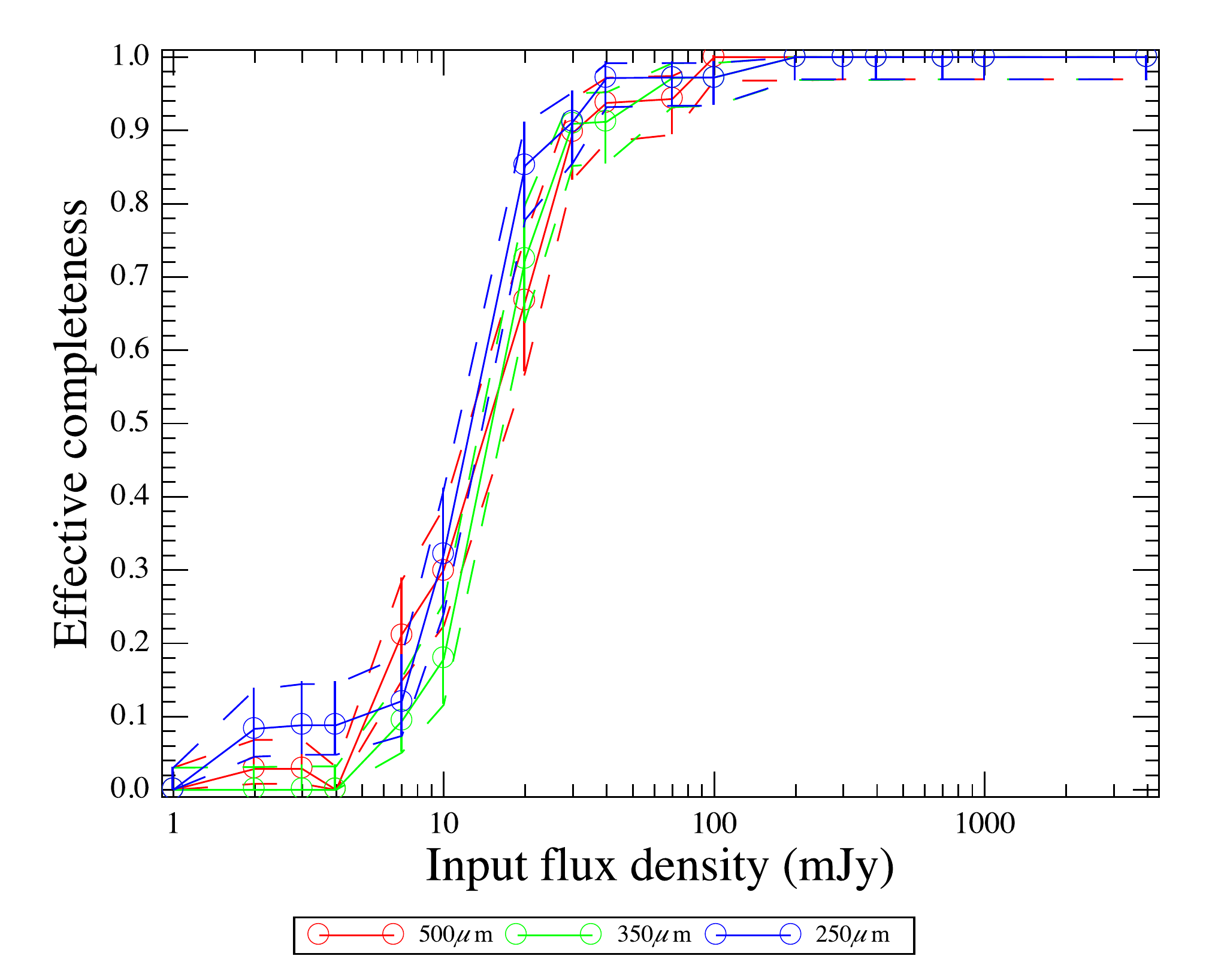}}
\subfigure[FLS]{\includegraphics[width=.45\textwidth]{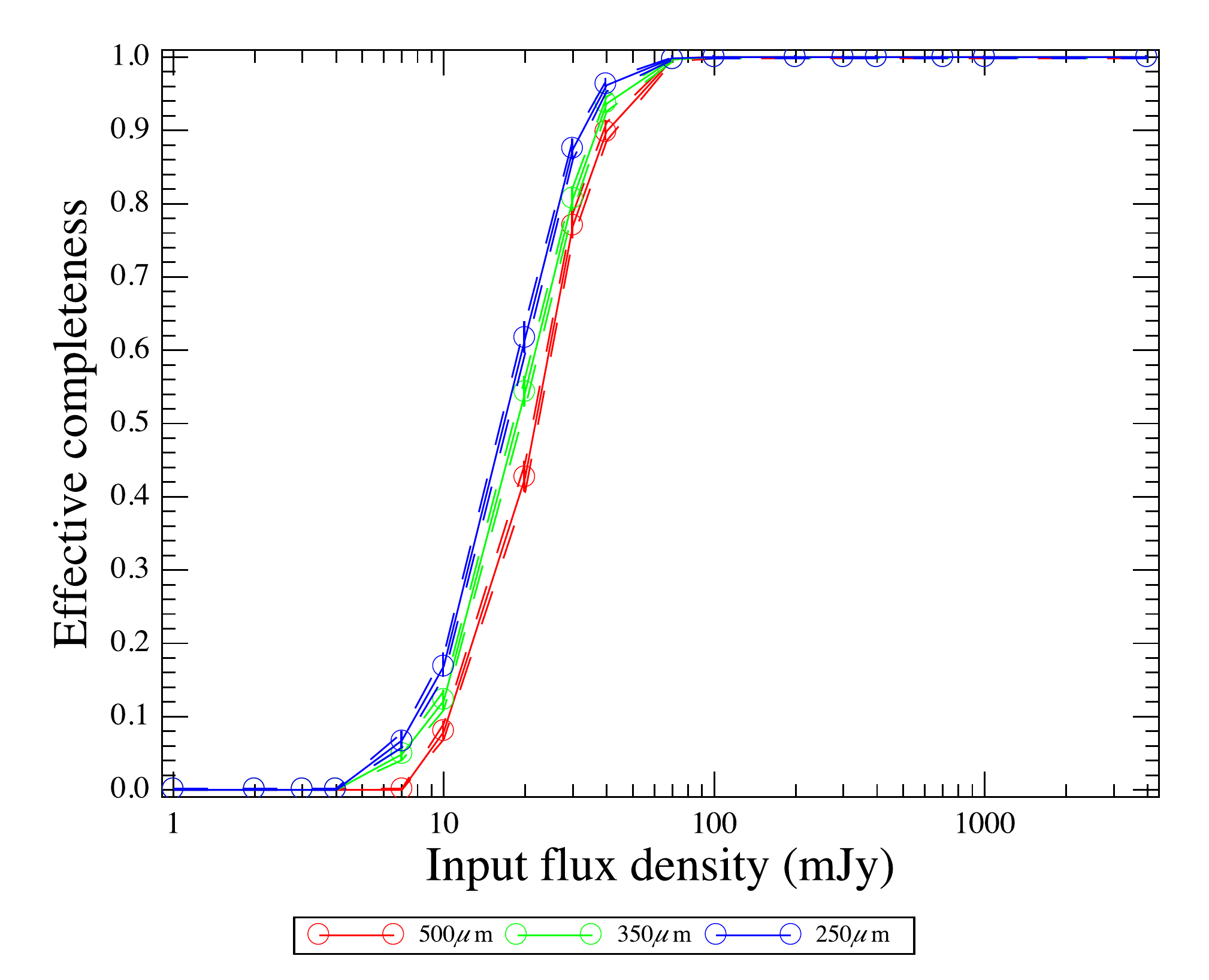}}
\subfigure[GOODS-North]{\includegraphics[width=.45\textwidth]{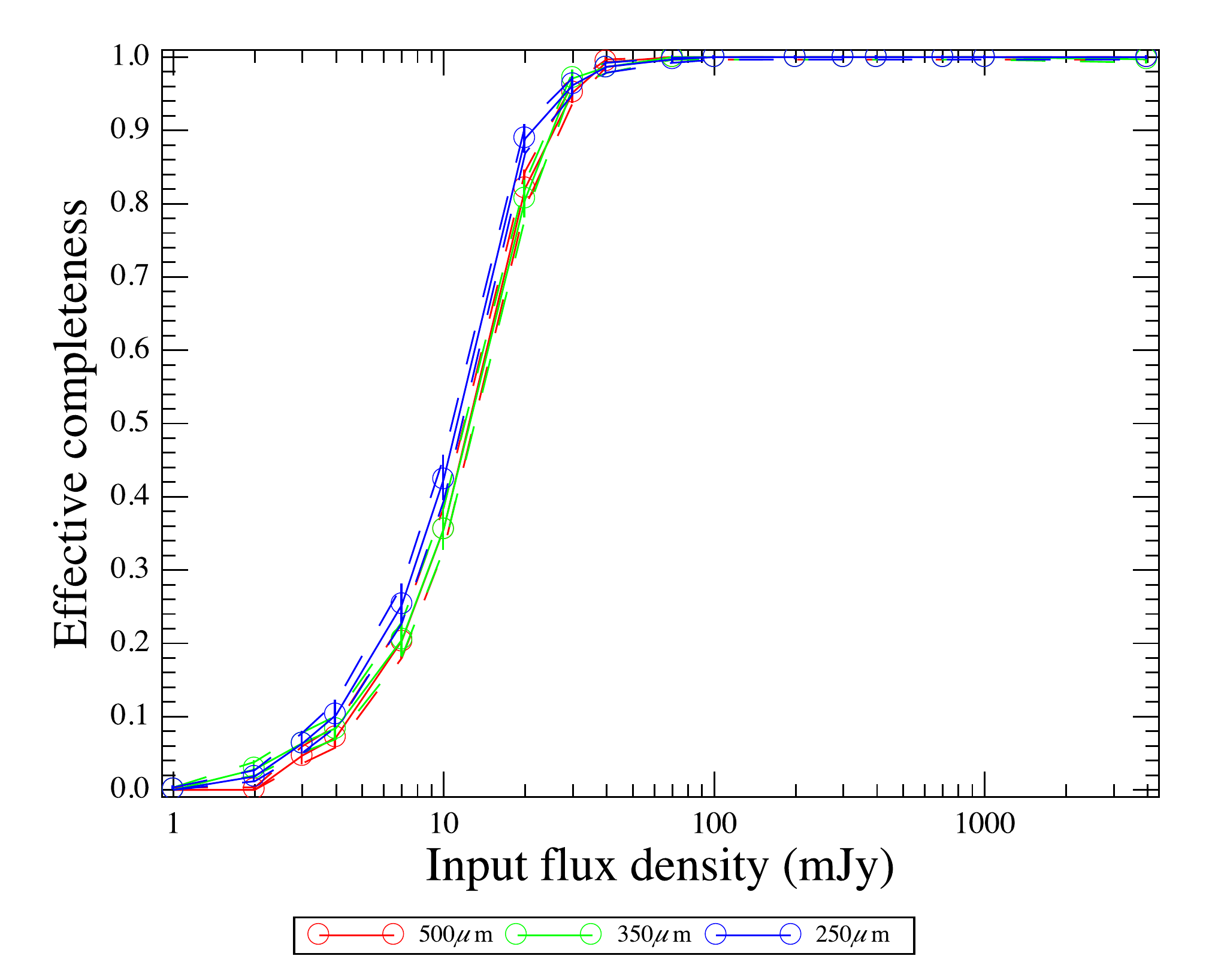}}
\subfigure[Lockman-North]{\includegraphics[width=.45\textwidth]{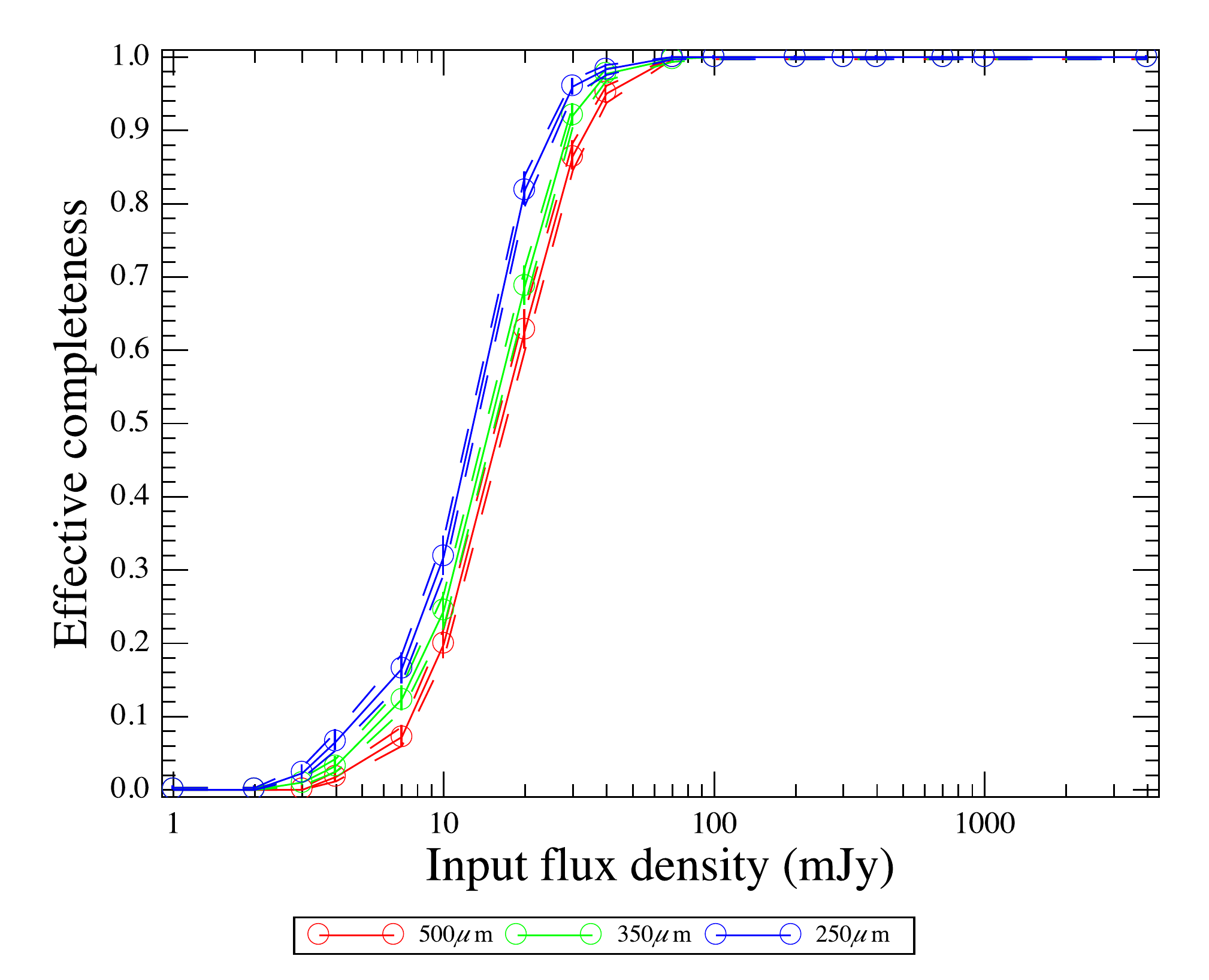}}
\subfigure[Lockman-SWIRE]{\includegraphics[width=.45\textwidth]{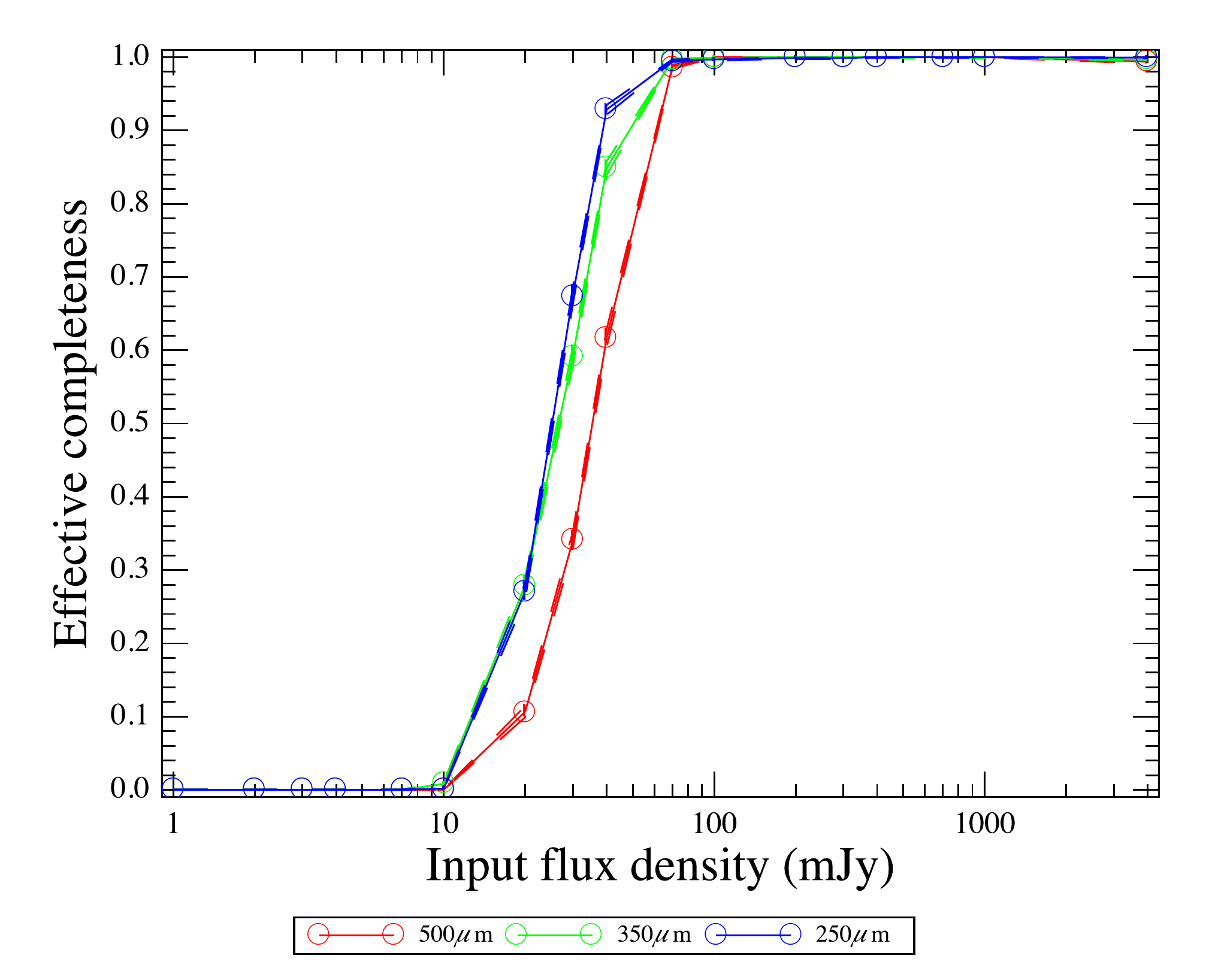}}
\caption{\label{fig:completeness}Completeness, for the five fields, as in Fig.\ \ref{fig:counts}. The flux density corresponding to 50 per cent completeness is given in Table \ref{tbl:completeness50}. Error bars and dashed lines are the lower and upper bounds of the posterior probability distribution for the completeness, chosen such that the probabilities at the bounds are equal, and so that the integrated probability between the bounds is 68 per cent.}
\end{figure*}

The completeness, as defined above, for each field is shown in Fig.\ \ref{fig:completeness}. Treating the completeness, $C$, as the parameter of a binomial distribution, the posterior probability for the value of the completeness being $C$ is given by a beta distribution:
\begin{eqnarray}
P(C | N_\mathrm{inj}, N_\mathrm{rec}) \propto C^{N_\mathrm{rec}} (1-C)^{N_\mathrm{inj} - N_\mathrm{rec}}
\end{eqnarray}
where $N_\mathrm{inj}$ is the number of injected sources and $N_\mathrm{rec}$ is the number of sources recovered. This is used to obtain the error bars on the completeness in Fig.\ \ref{fig:completeness}.

For each completeness curve, the flux density is found at which the completeness is 50 per cent; these values are given in Table \ref{tbl:completeness50}. 

\begin{table}
\caption{\label{tbl:completeness50}Flux density corresponding to 50 per cent completeness in Fig.\ \ref{fig:completeness}.}
\vspace{0.2cm}
\centering
\begin{tabular}{l|rrr}
\hline
\hline
Field 	& \multicolumn{3}{c}{Flux density, mJy} \\
	 	& 250\,$\mu$m 	& 350\,$\mu$m 	& 500\,$\mu$m \\
\hline
A2218 			& 13.4 	& 15.9 	& 15.5 \\
FLS	    			& 17.4	& 19.0 	& 22.1 \\
GOODS-North		& 11.6 	& 13.2	& 13.1 \\
Lockman-North		& 13.6 	& 15.7	& 17.0 \\
Lockman-SWIRE	& 25.7 	& 27.1	& 35.8 \\
\hline
\hline
\end{tabular}
\end{table}

\begin{figure*}
\centering
\subfigure[A2218]{\includegraphics[width=.45\textwidth]{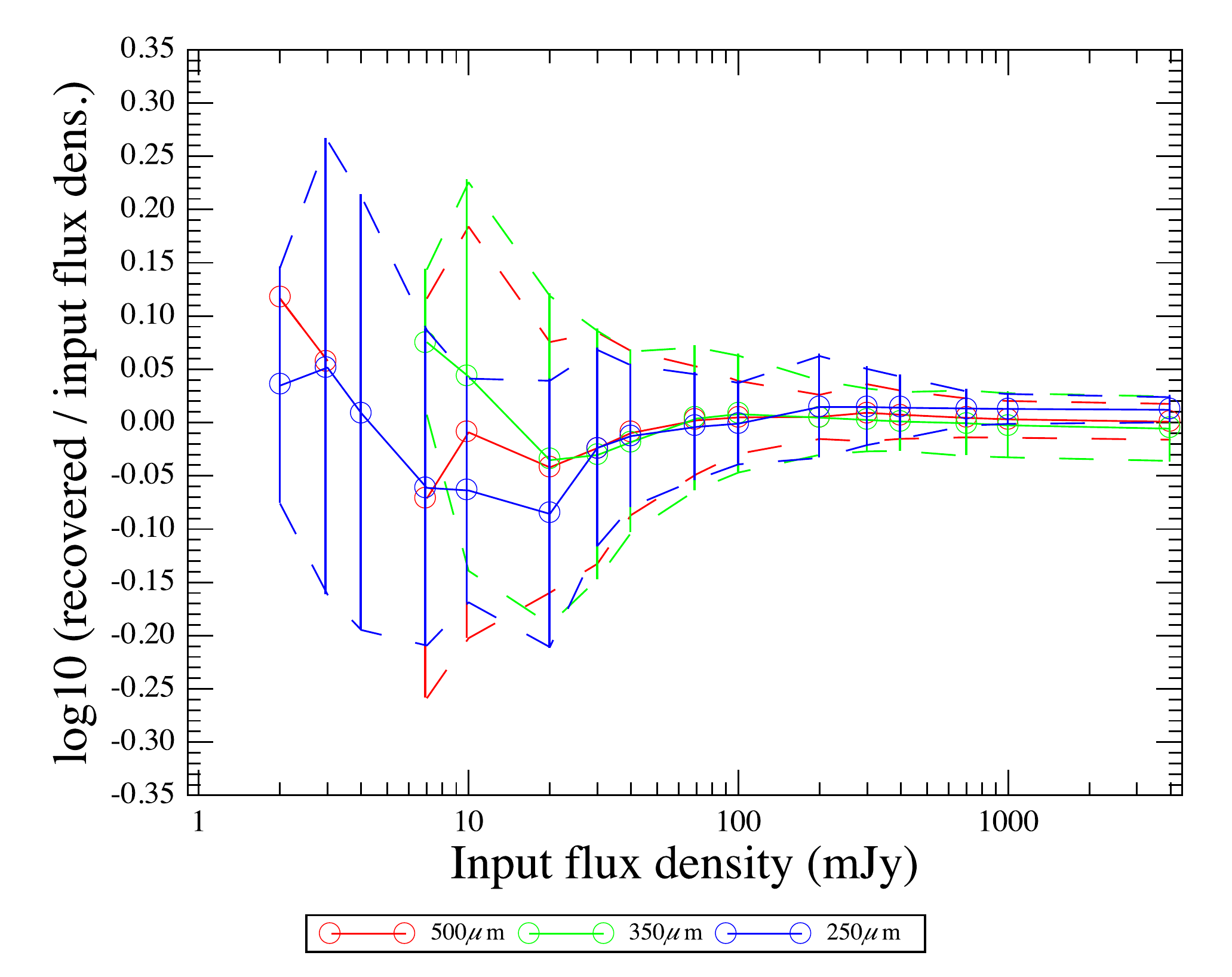}}
\subfigure[FLS]{\includegraphics[width=.45\textwidth]{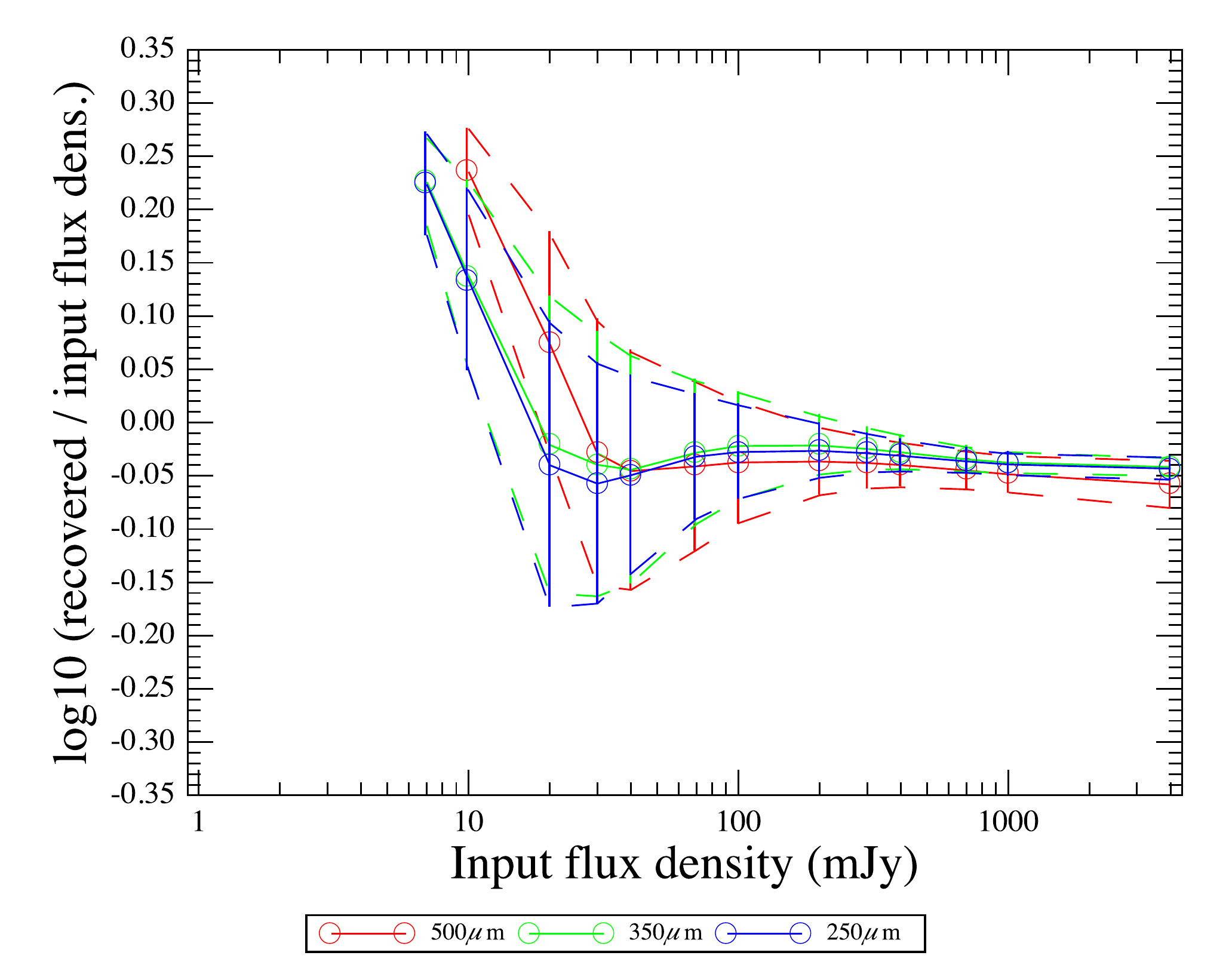}}
\subfigure[GOODS-North]{\includegraphics[width=.45\textwidth]{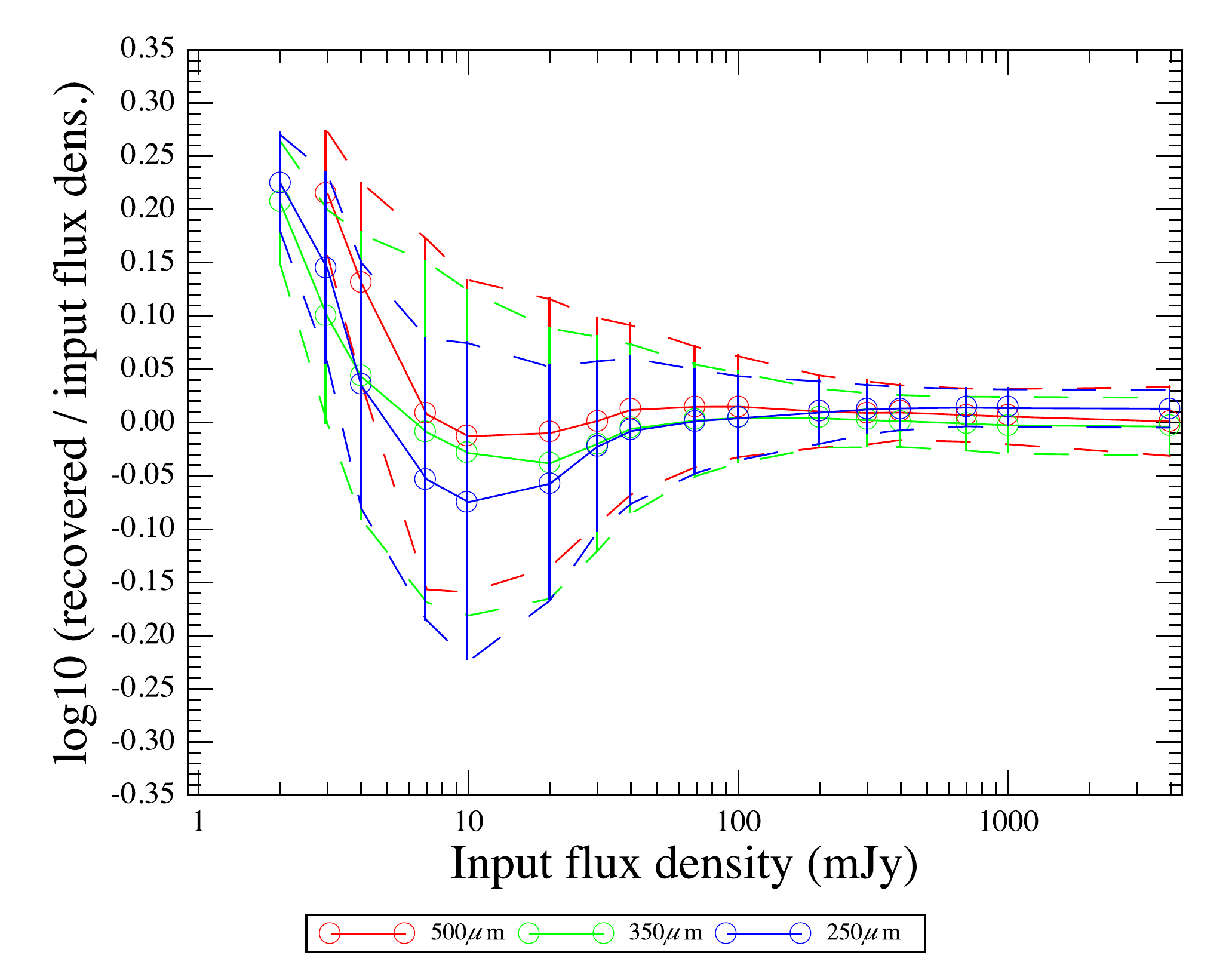}}
\subfigure[Lockman-North]{\includegraphics[width=.45\textwidth]{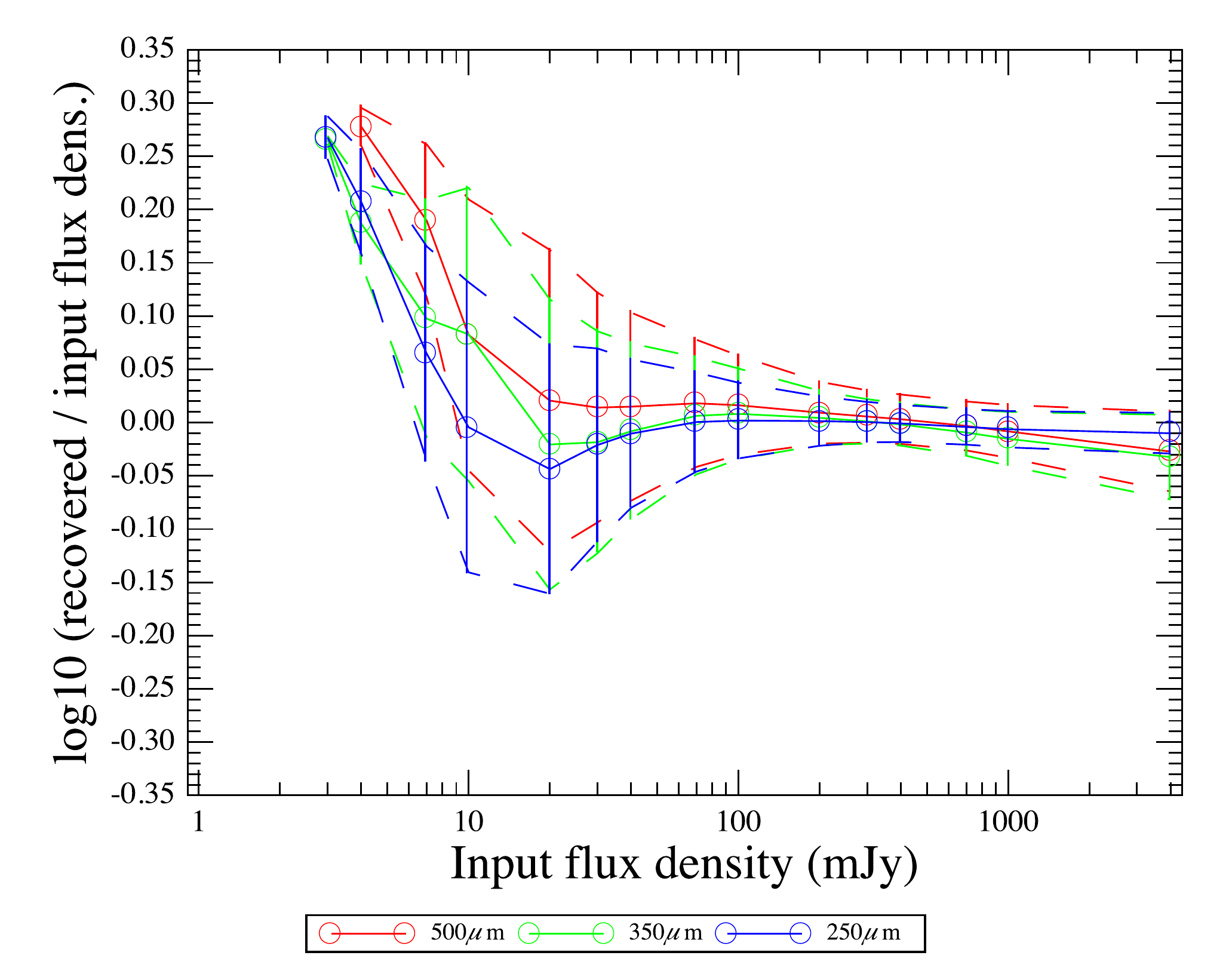}}
\subfigure[Lockman-SWIRE]{\includegraphics[width=.45\textwidth]{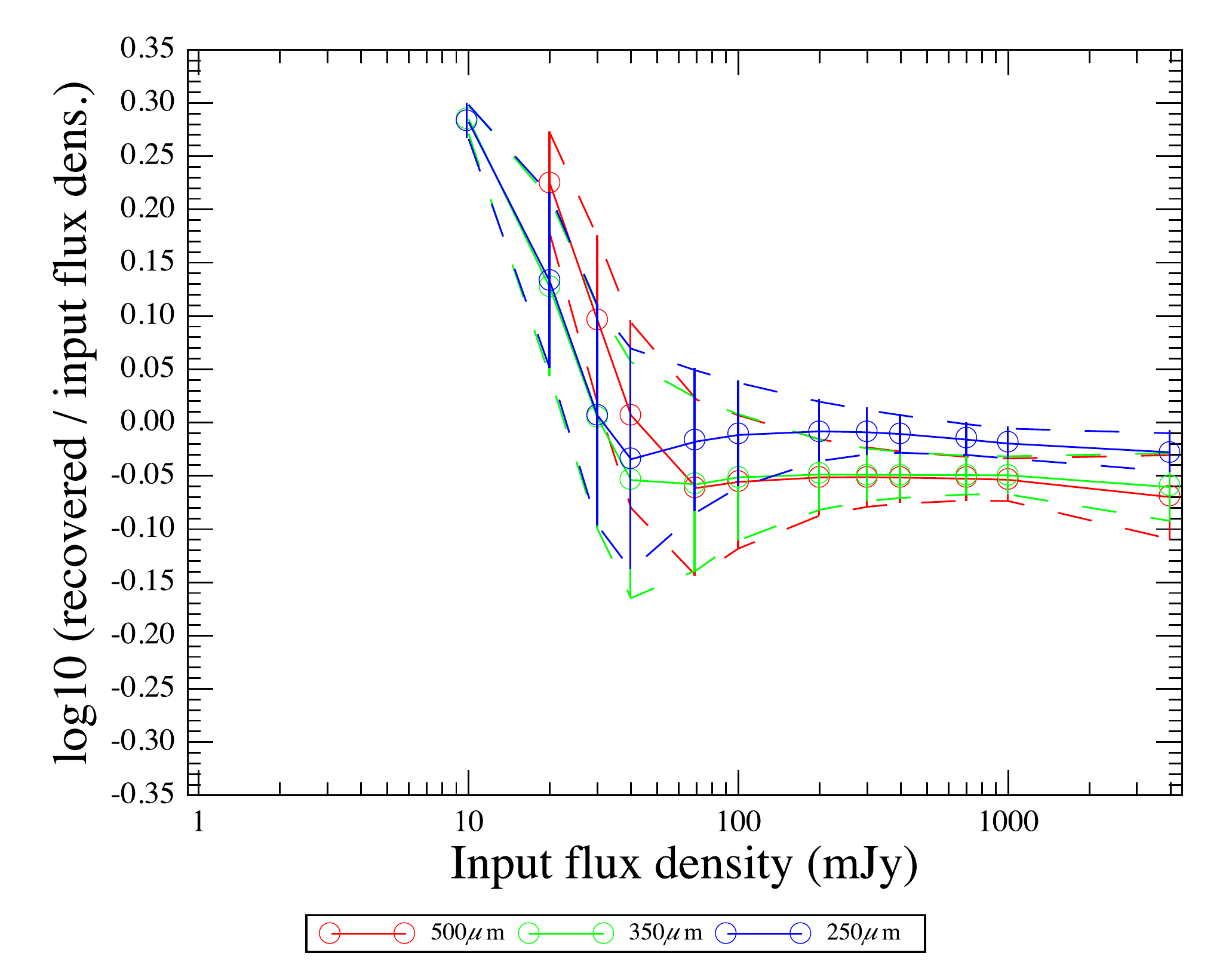}}
\caption{\label{fig:flux}Flux density accuracy, for the five fields, as in Fig.\ \ref{fig:counts}. Error bars are the RMS of the log$_{10}$ (recovered flux density $/$ input flux density) at that flux density.}
\end{figure*}

\begin{figure*}
\centering
\subfigure[A2218]{\includegraphics[width=.45\textwidth]{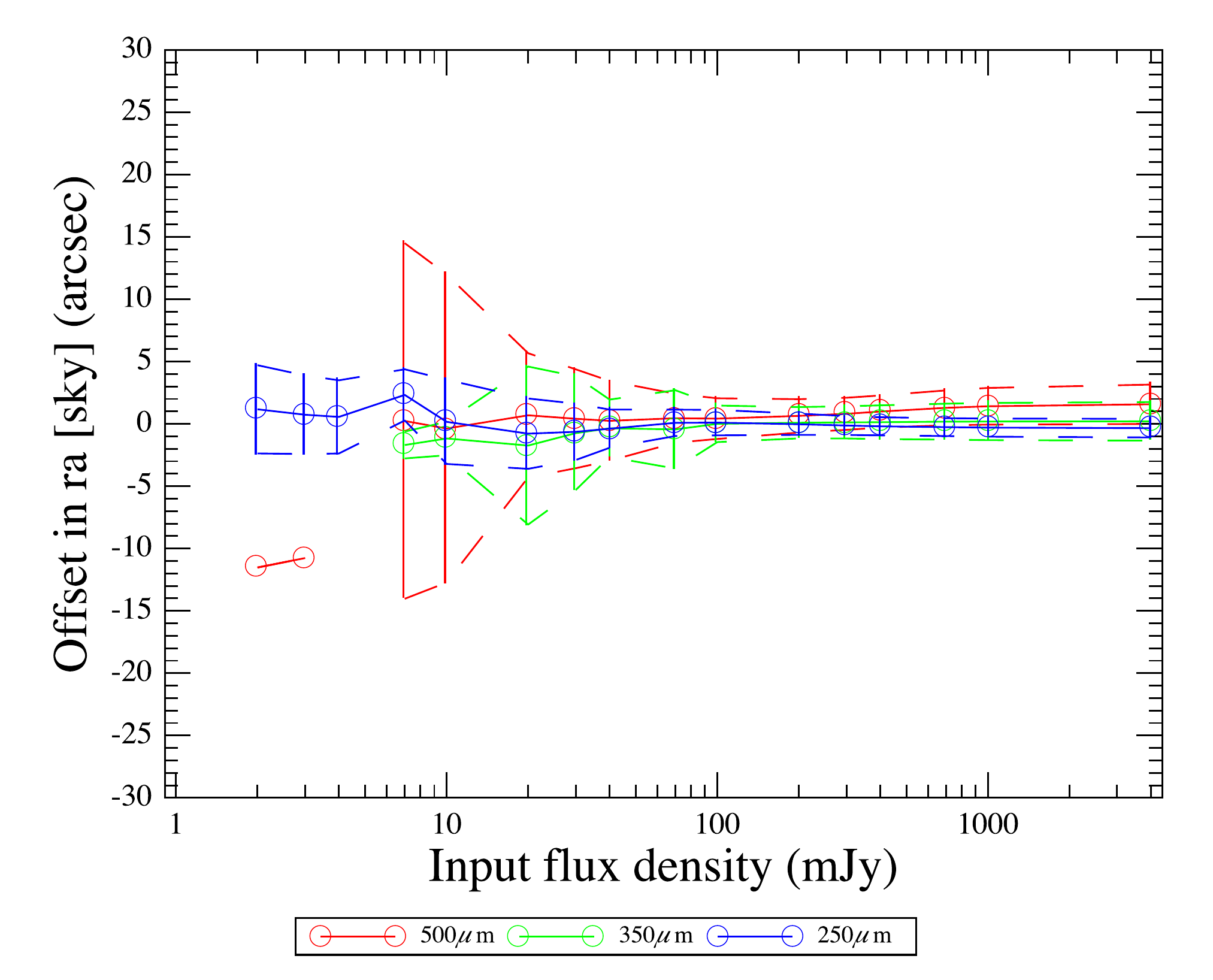}}
\subfigure[FLS]{\includegraphics[width=.45\textwidth]{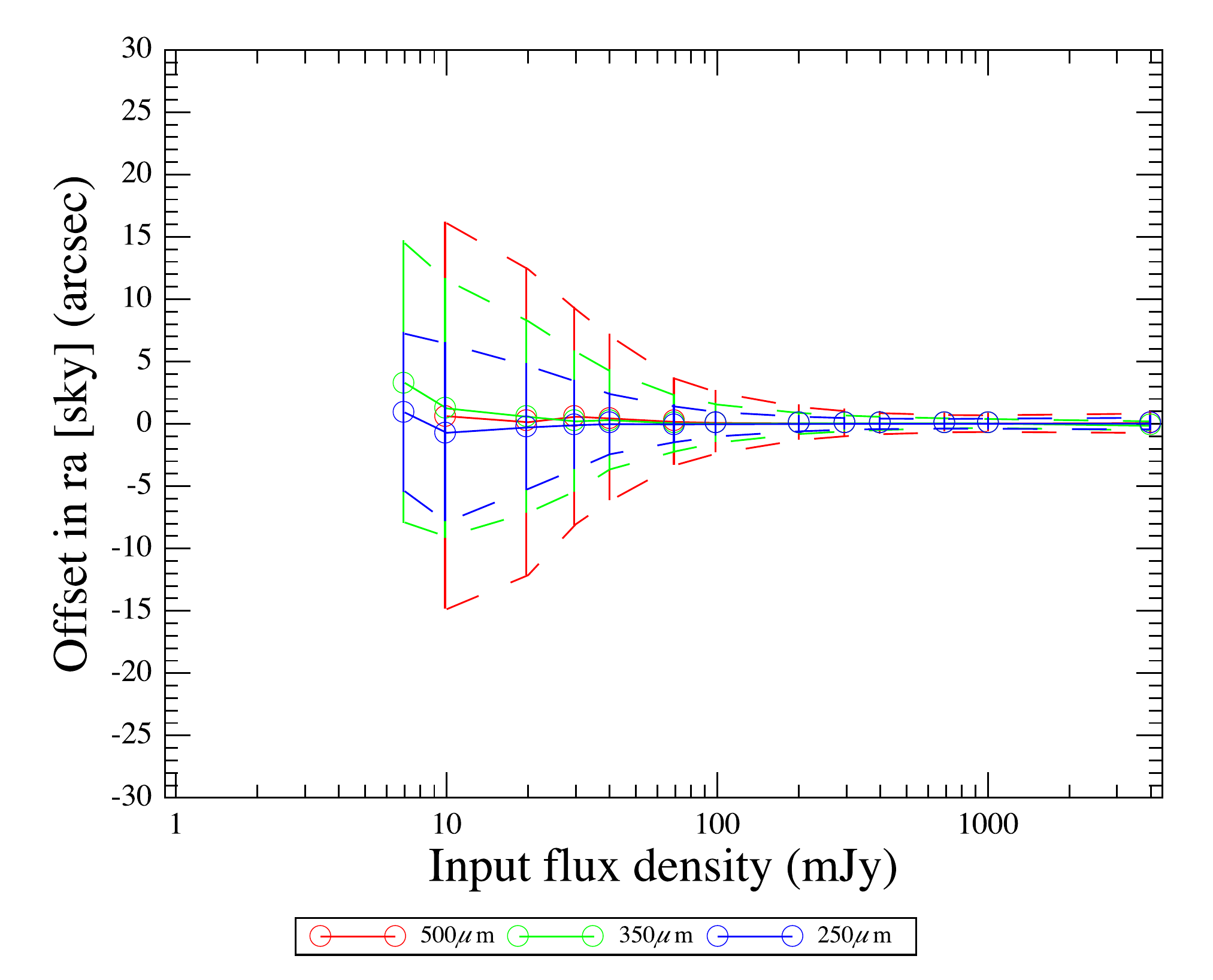}}
\subfigure[GOODS-North]{\includegraphics[width=.45\textwidth]{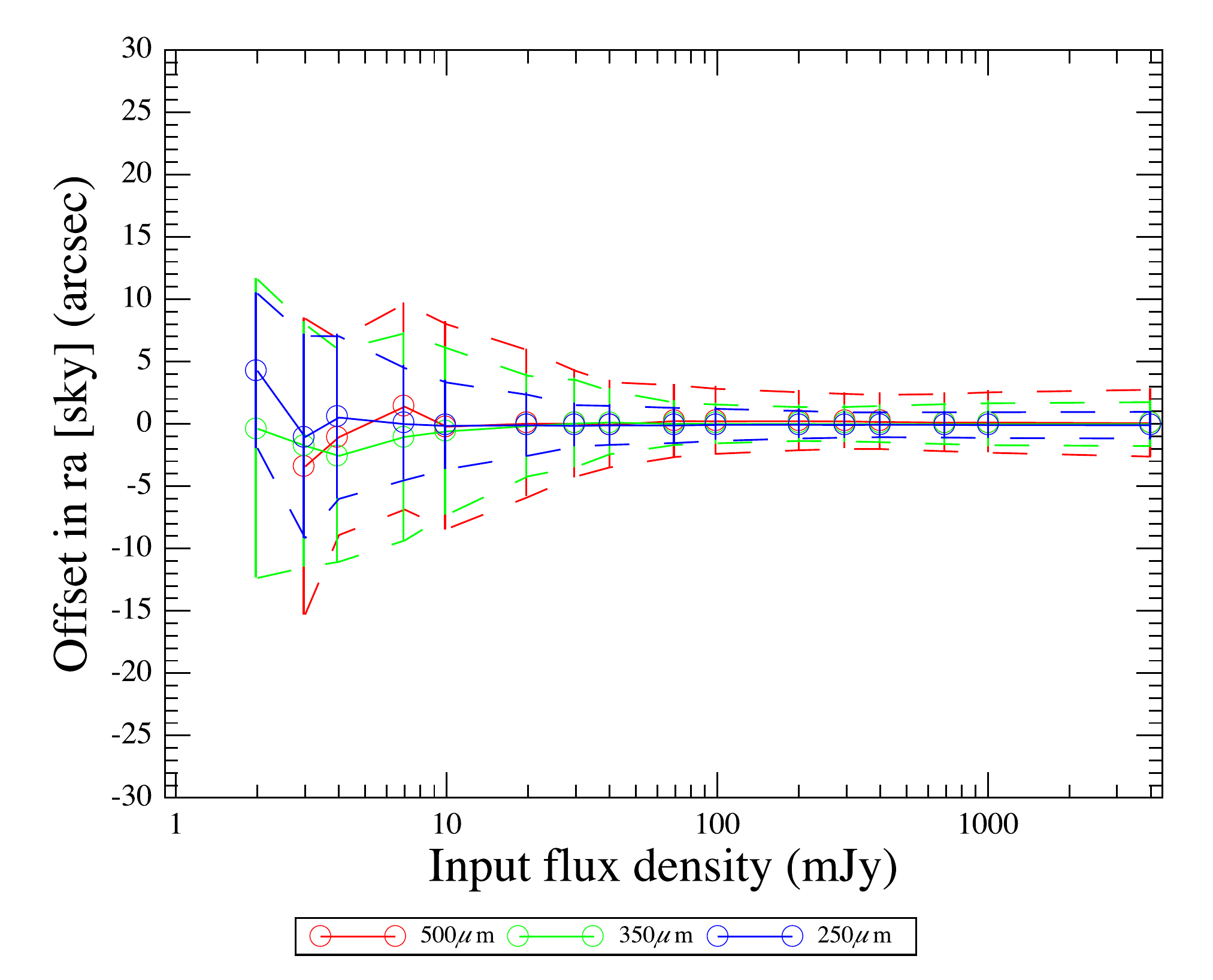}}
\subfigure[Lockman-North]{\includegraphics[width=.45\textwidth]{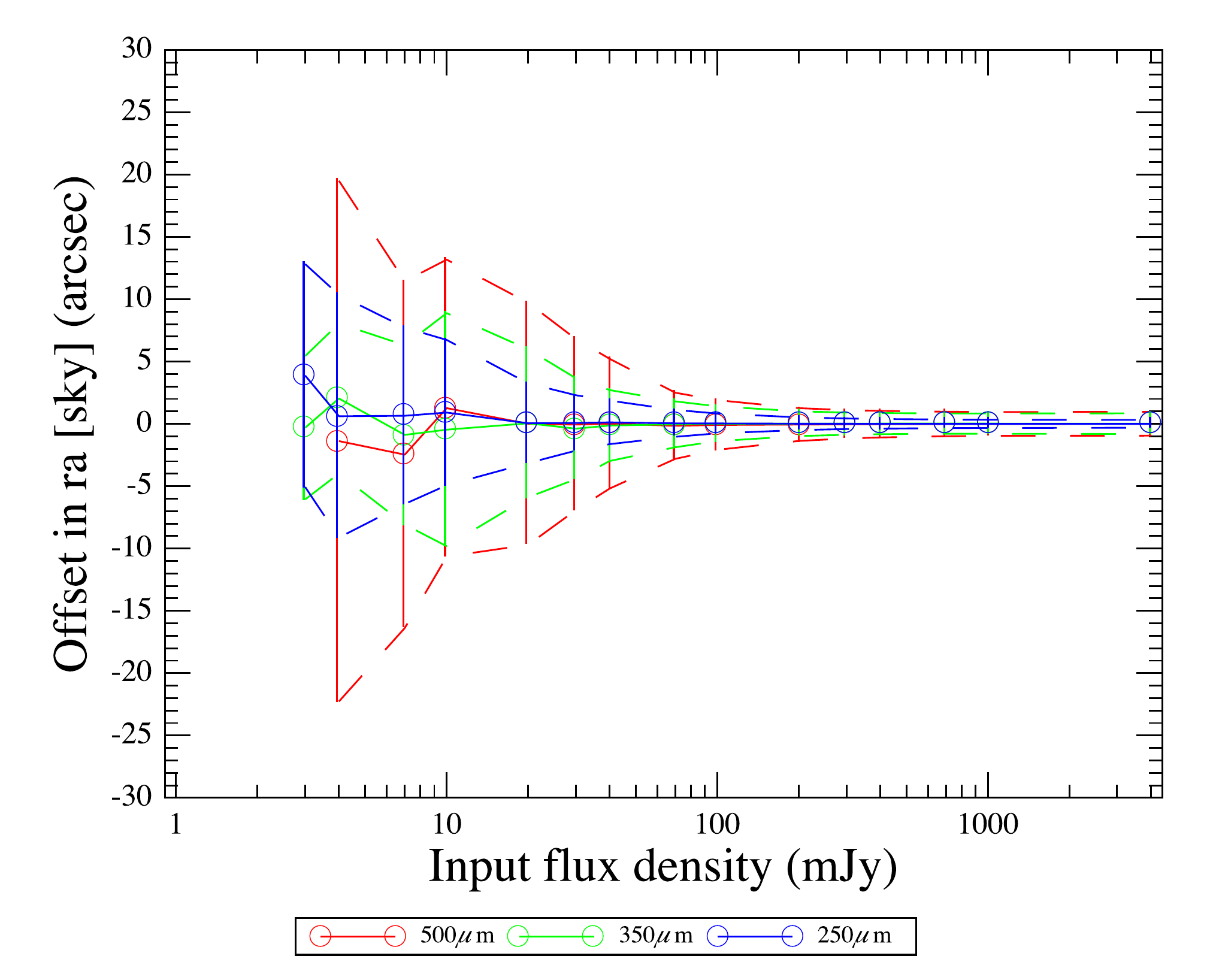}}
\subfigure[Lockman-SWIRE]{\includegraphics[width=.45\textwidth]{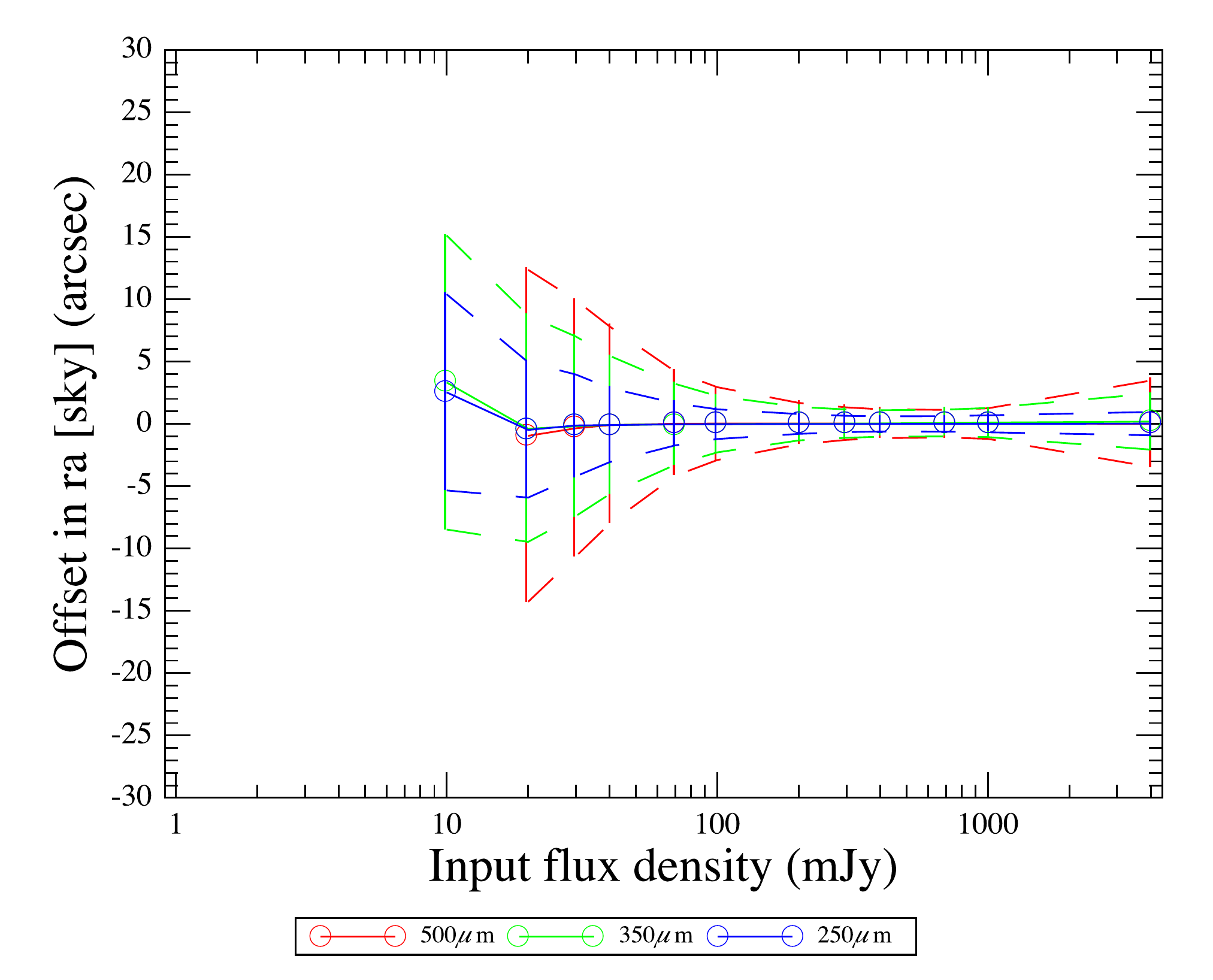}}
\caption{\label{fig:ra}Positional accuracy in RA (offset to the east, in arcsec), for the five fields, as in Fig.\ \ref{fig:counts}. Error bars are the RMS of the positional error at that flux density. Similar results are found for the offset in declination.}
\end{figure*}

The accuracy of the flux densities and positions of the recovered sources are shown in Figs.\ \ref{fig:flux} and \ref{fig:ra} respectively.

Table \ref{tbl:flux_mean} shows the mean offset in flux density for various input flux densities, while Table \ref{tbl:flux_rms} shows the corresponding RMS scatter in the flux density offset and Table \ref{tbl:ra_rms} shows the RMS scatter for the offset in RA.

\begin{table*}
\caption{\label{tbl:flux_mean}Mean value of log$_{10}$ (recovered flux density $/$ input flux density), for each band, at input flux densities of 20, 40, 100 and 1000\,mJy, from Fig.\ \ref{fig:flux}.}
\vspace{0.2cm}
\centering
\begin{tabular}{lrrrrrrrrrrrr}
\hline
\hline
Field 	& \multicolumn{12}{c}{Mean offset, dex} \\
	 	& \multicolumn{4}{c}{250\,$\mu$m} 	& \multicolumn{4}{c}{350\,$\mu$m} 	& \multicolumn{4}{c}{500\,$\mu$m} \\
	 	& 20 & 40 & 100 & 1000 & 20 & 40 & 100 & 1000 & 20 & 40 & 100 & 1000\,mJy \\
\hline
A2218	 & -0.086 & -0.013 & -0.001 & 0.013 	 & -0.035 & -0.018 & 0.008 & -0.003 	 & -0.042 & -0.010 & 0.005 & 0.003 \\
FLS	 & -0.040 & -0.049 & -0.028 & -0.039 	 & -0.021 & -0.044 & -0.022 & -0.038 	 & 0.074 & -0.046 & -0.037 & -0.049 \\
GOODSN	 & -0.057 & -0.008 & 0.004 & 0.013 	 & -0.038 & -0.006 & 0.004 & -0.003 	 & -0.010 & 0.012 & 0.015 & 0.006 \\
LOCKN	 & -0.044 & -0.010 & 0.002 & -0.006 	 & -0.021 & -0.008 & 0.008 & -0.015 	 & 0.021 & 0.015 & 0.016 & -0.008 \\
LOCKSW	 & 0.132 & -0.034 & -0.012 & -0.020 	 & 0.126 & -0.054 & -0.051 & -0.050 	 & 0.225 & 0.007 & -0.056 & -0.054 \\
\hline
\hline
\end{tabular}
\end{table*}

\begin{table*}
\caption{\label{tbl:flux_rms}RMS scatter in log$_{10}$ (recovered flux density $/$ input flux density), for each band, at input flux densities of 20, 40, 100 and 1000\,mJy, from Fig.\ \ref{fig:flux}.}
\vspace{0.2cm}
\centering
\begin{tabular}{lrrrrrrrrrrrr}
\hline
\hline
Field 	& \multicolumn{12}{c}{RMS scatter, dex} \\
	 	& \multicolumn{4}{c}{250\,$\mu$m} 	& \multicolumn{4}{c}{350\,$\mu$m} 	& \multicolumn{4}{c}{500\,$\mu$m} \\
	 	& 20 & 40 & 100 & 1000 & 20 & 40 & 100 & 1000 & 20 & 40 & 100 & 1000\,mJy \\
\hline
A2218	 & 0.125 & 0.066 & 0.038 & 0.014 	 & 0.154 & 0.085 & 0.055 & 0.030 	 & 0.118 & 0.077 & 0.034 & 0.017 \\
FLS	 & 0.134 & 0.092 & 0.044 & 0.010 	 & 0.139 & 0.107 & 0.050 & 0.010 	 & 0.103 & 0.112 & 0.057 & 0.017 \\
GOODS-North	 & 0.110 & 0.068 & 0.039 & 0.018 	 & 0.127 & 0.079 & 0.042 & 0.027 	 & 0.126 & 0.079 & 0.047 & 0.026 \\
LOCK-North	 & 0.117 & 0.070 & 0.036 & 0.017 	 & 0.136 & 0.083 & 0.043 & 0.025 	 & 0.141 & 0.088 & 0.047 & 0.025 \\
LOCK-SWIRE	 & 0.082 & 0.104 & 0.049 & 0.014 	 & 0.083 & 0.111 & 0.059 & 0.018 	 & 0.047 & 0.087 & 0.063 & 0.020 \\
\hline
\hline
\end{tabular}
\end{table*}

\begin{table*}
\caption{\label{tbl:ra_rms}RMS scatter in the offset in RA, for each band, at input flux densities of 20, 40, 100 and 1000\,mJy, from Fig.\ \ref{fig:ra}.}
\vspace{0.2cm}
\centering
\begin{tabular}{lrrrrrrrrrrrr}
\hline
\hline
Field 	& \multicolumn{12}{c}{RMS scatter, arcsec} \\
	 	& \multicolumn{4}{c}{250\,$\mu$m} 	& \multicolumn{4}{c}{350\,$\mu$m} 	& \multicolumn{4}{c}{500\,$\mu$m} \\
	 	& 20 & 40 & 100 & 1000 & 20 & 40 & 100 & 1000 & 20 & 40 & 100 & 1000\,mJy \\
\hline
A2218	 & 2.821 & 1.533 & 1.019 & 0.727 	 & 6.337 & 2.301 & 1.466 & 1.491 	 & 4.980 & 3.200 & 1.631 & 1.472  \\
FLS	 & 4.935 & 2.418 & 0.976 & 0.403 	 & 7.679 & 3.954 & 1.528 & 0.360 	 & 12.272 & 6.628 & 2.440 & 0.668 \\
GOODS-North	 & 2.439 & 1.564 & 1.283 & 1.035 	 & 4.030 & 2.546 & 1.554 & 1.681 	 & 5.883 & 3.418 & 2.604 & 2.420 \\
LOCK-North	 & 3.196 & 1.746 & 0.795 & 0.328 	 & 5.935 & 2.846 & 1.396 & 0.815 	 & 9.707 & 5.206 & 1.993 & 0.958 \\
LOCK-SWIRE	 & 5.428 & 2.971 & 1.190 & 0.681 	 & 9.077 & 5.552 & 2.258 & 1.166 	 & 13.314 & 7.916 & 2.922 & 1.226 \\
\hline
\hline
\end{tabular}
\end{table*}

Several points should be noted from these.
\begin{enumerate}
\item The source extraction method has been adjusted by applying a multiplicative factor to all flux densities in order to give good recovered flux densities for bright injected sources. These factors are given in Section \ref{sec:catalogues}. This can be seen in Fig.\ \ref{fig:flux} by the way the flux density offset is measured to be approximately zero for bright injected sources for most of the fields (some late adjustments were made to the Wiener filtered data, leaving a very small residual offset for bright flux densities for FLS and Lockman-SWIRE).
\item In Fig.\ \ref{fig:flux}, towards fainter flux densities, there is a trend of a bias towards an overestimated flux density and decreasing RMS scatter; this is a selection effect due to flux boosting and the requirement that a good match will have a flux density within a factor of 2 (0.3 dex) of the input flux density.
\item Also in Fig.\ \ref{fig:flux} there is a dip in the plots at around 10--40\,mJy, corresponding to an underestimate of the flux density. This is likely to be due to the maps having a zero mean, leading to a systematic underestimation of the flux densities. Corrective factors were applied to the flux densities in order to give good agreement for bright input sources, but these were multiplicative corrections, rather than corrections with both an additive and multiplicative component, and thus the corrections have been effective only for bright input sources, where the multiplicative factor is dominant.
\item The scatter in the recovered flux densities in Fig.\ \ref{fig:flux} may be compared with the uncertainties in the flux densities given in the catalogues. The latter are, to first order, independent of the flux density of the source, being based on the statistics of the whole map (see Section \ref{sec:uncertainties}). However, for bright injected sources, the former is approximately proportional to the input flux density. This is because the dominant source of noise in the measurement of the flux density of a bright source is not confusion noise but rather systematic errors from the source extraction method, for example, variations in the recovered flux density depending on where the centre of a source lies within a pixel (see Section \ref{sec:catalogues}). For bright sources, therefore, the values given in Table \ref{tbl:flux_rms} should be used as an approximate guide to the uncertainty in the measured flux.
\item The quality of the positions and flux densities for the brightest flux densities can be seen from Tables \ref{tbl:flux_rms} and \ref{tbl:ra_rms} to depend on the approach used to extract the sources. This is discussed in Section \ref{sec:catalogues} and depends on whether the shallow approach has been used (FLS, Lockman-North and Lockman-SWIRE), or the deep approach (A2218 and GOODS-North) and whether a Wiener filter has been applied to the maps (FLS and Lockman-SWIRE). In particular, where a smaller amount of smoothing has been applied to the maps in order to improve the extraction of faint sources and deal with the problem of source blending (A2218 and GOODS-North), the scatter in the recovered flux densities and positions for bright injected sources is larger than when more smoothing is applied.
\item The combined effect of the flux density uncertainties in Fig.\ \ref{fig:flux} and the steep number counts seen at SPIRE wavelengths \citep{Oliver...2010} will mean that at any given measured flux density, most of the sources will have a true flux density which is fainter than the measured flux density, even if there is no systematic offset of injected to measured flux densities, as is the case here. This phenomenon (`flux boosting') must be taken into account when estimating the true flux densities of sources. See \citet{Oliver...2010} for further discussion.
\item Some features may be discerned for bright injected flux densities in Fig.\ \ref{fig:flux}, such as a departure away from a horizontal slope. This is due to the effect of the injected sources on the error maps. In the steep slope of the point-response function, there is a large scatter in the intensity of the bolometer samples falling within the map pixel. This leads through Equation (\ref{mapfluxerror}) to a higher value for the uncertainty for that map pixel, which leads through Equation (\ref{eqn:fluxerror}) to that map pixel being given a lower weight in the source extraction. These changes in the relative weight given to the pixels in a point source lead to changing estimates of the source flux density, based on the flux density of the injected source.
\item Finally, it should be noted that the method used here to evaluate the accuracy of the measured flux densities and positions uses idealised artificial sources, which will be subtly different from real sources. One cause of these differences would be the assumed Gaussian PRF; the Airy rings around an extremely bright source would influence the detections close to that source. For the very brightest sources, the best way to test \textsc{sussextractor} (and other algorithms) would be on real observations of SPIRE calibration sources \citep[see][]{Swinyard...2010a}.
\end{enumerate}

\section{Conclusions}
\label{sec:conclusions}

We have described the approach adopted to generate single-band catalogues from HerMES SDP SPIRE observations, some of which have been made publicly available. A formalism has been developed to assess the quality of these catalogues, and recommendations have been made for usage of the catalogues based on these results.

Possible improvements to the method used here have been identified above, including the following:
\begin{enumerate}
\item The filtering of the data could be refined, both by including information about cirrus in the noise spectrum for the Wiener filter (Section \ref{sec:maps}), and by using optimized matched filters for the detection and measurement of point sources (Section \ref{sec:catalogues}).
\item A better estimation of the background would deal with the additive offset to the measured flux densities, discussed in Section \ref{sec:completeness_results}. This could be achieved either by determining a physical zero-point for the maps or through local background estimation for each source.
\item A better estimation of the flux densities could be achieved by relaxing the implicit assumption that all sources lie in the centre of map pixels. This could be achieved by first finding the source positions to sub-pixel accuracy, as in the current method, and then choosing an appropriate smoothing kernel in Equation (\ref{eqn:flux}) based on this measured position.
\end{enumerate}

Only one approach has been presented here for source extraction. Other approaches exist, both in terms of other algorithms, and alternative ways of using any particular algorithm, such as an iterative approach, removing the brightest sources from the image at each iteration. Alternatively, information from multiple bands may be used, either simultaneously, to extract sources in multiple \textit{Herschel} bands at the same time, or by using prior positions from other wavelengths, a method used for many HerMES results and described by \citet{Roseboom...2010}.  Future work within HerMES will explore these approaches in more depth.

\section*{Acknowledgments}

SPIRE has been developed by a consortium of institutes led by Cardiff Univ. (UK) and including Univ. Lethbridge (Canada); NAOC (China); CEA, LAM (France); IFSI, Univ. Padua (Italy); IAC (Spain); Stockholm Observatory (Sweden); Imperial College London, RAL, UCL-MSSL, UKATC, Univ. Sussex (UK); Caltech, JPL, NHSC, Univ. Colorado (USA). This development has been supported by national funding agencies: CSA (Canada); NAOC (China); CEA, CNES, CNRS (France); ASI (Italy); MCINN (Spain); SNSB (Sweden); STFC (UK); and NASA (USA).

The released HerMES data are available through the HeDaM database (\url{http://hedam.oamp.fr}), operated by CeSAM and hosted by the Laboratoire d'Astrophysique de Marseille.

Data presented in this paper were analysed using \textsc{hipe}, a joint development by the \textit{Herschel} Science Ground Segment Consortium, consisting of ESA, the NASA \textit{Herschel} Science Center, and the HIFI, PACS and SPIRE consortia. (See \url{http://herschel.esac.esa.int/DpHipeContributors.shtml}.)

We acknowledge support from the Science and Technology Facilities Council [grant number ST/F002858/1] and [grant number ST/I000976/1].

We are grateful to the anonymous referee for helpful and constructive comments.


\appendix
\section{Data release}
\label{sec:release}

The central region of the maps for A2218 and part of the catalogues for A2218, Lockman-SWIRE and FLS described in this paper have been released to the public on the Herschel Database in Marseille (HeDaM).\footnote{\url{http://hedam.oamp.fr/HerMES/}}

For A2218, three single-band SPIRE catalogues have been released, including all sources within the central region with flux densities greater than 20\,mJy (6 sources at 500\,$\mu$m, 19 at at 350\,$\mu$m and 35 at 250\,$\mu$m). Descriptions of the columns are given in Table \ref{tbl:columns}.

For FLS and Lockman-SWIRE, 250\,$\mu$m catalogues have been released, containing all sources within the central region with flux densities greater than 100\,mJy, as long as the source has one and only one counterpart within 10 arcsec in the associated 24\,$\mu$m catalogue. The resulting FLS catalogue contains 45 sources and the Lockman-SWIRE catalogues contains 114. For FLS and Lockman-SWIRE, in addition to the columns described in Table \ref{tbl:columns}, additional columns are provided in the catalogues, some derived from a fusion of ancillary data (Vaccari et al., in preparation) and others containing SPIRE list-driven quantities derived from 24\,$\mu$m source positions \citet{Roseboom...2010}.

\begin{table*}
\caption{\label{tbl:columns}Names and descriptions of the columns included in the public data release (see Appendix \ref{sec:release}).}
\vspace{0.2cm}
\centering
\begin{tabular}{l r p{12cm}}
\hline
\hline
Name & Col.\ No. & Description\\
\hline
name & 1 & HerMES ID \\
ra & 2 & Right Ascension (deg)\\
dec & 3 & Declination (deg)\\
raErr & 4 & Right ascension uncertainty (deg)\\
decErr & 5 & Declination uncertainty (deg)\\
flux & 6 & Source flux density (mJy)\\
fluxErr & 7 & Source flux density formal uncertainty (instrumental noise, mJy)\\
quality & 8 & Signal to instrumental noise: flux/fluxErr\\
index & 9 & Sequential number in full catalogue\\
centralRegion & 10 & True if source lies within a well defined central region of the map\\
fluxErrTotal & 11 & Total uncertainty in the source flux density, due to confusion and instrumental noise (mJy)\\
SNR & 12 & Signal to total noise: flux/fluxErrTotal\\
fluxHalfData1 & 13 & Source flux density, as measured using a map based on the first half of the data (mJy) \\
qualityHalfData1 & 14 & Signal to instrumental noise, as measured using a map based on the first half of the data\\
fluxHalfData2 & 15 & Source flux density, as measured using a map based on the second half of the data (mJy) \\
qualityHalfData2 & 16 & Signal to instrumental noise, as measured using a map based on the second half of the data\\
extended & 177/178 & If true, source noticeably extended in the SPIRE 250\,$\mu$m image (flux density should be treated with caution) [not A2218]\\
\hline
\hline
\end{tabular}
\end{table*}

\label{lastpage}

\end{document}